\documentclass[12pt]{article}
\usepackage{amssymb}
\usepackage{natbib,graphicx,setspace,lscape,longtable}
\usepackage{natbib,epsfig,graphicx}
\usepackage{amsmath,amsthm,amssymb,color}
\usepackage{enumitem}
\usepackage{mathtools}
\usepackage{array}
\usepackage{booktabs}
\usepackage{url}
\usepackage{ulem}
\usepackage{subcaption} 
\usepackage{amsfonts,amssymb}
\usepackage{hhline}
\usepackage{mathrsfs}
\usepackage{natbib,graphicx,setspace,lscape,longtable,epstopdf,xr}
\usepackage{amsmath,amsthm,color}
\usepackage{natbib,epsfig,graphicx,pdfpages}
\usepackage{rotating}
\usepackage{float}
\usepackage{bm}
\usepackage{ragged2e}
\usepackage{todonotes}
\usepackage{hyperref}
\bibpunct{(}{)}{;}{a}{,}{,}

\newtheorem{theorem}{Theorem}
\newtheorem{lemma}{Lemma}
\newcommand{\csection}[1]
    {\begin{center}
        \stepcounter{section}
        {\bf\large\arabic{section}. #1}
    \end{center}
    \vspace{-0.15 cm}
}

\newcommand{\scsection}[1]
    {\begin{center}
        {\bf\large #1}
    \end{center}
    \vspace{-0.15 cm}
}

\newcommand{\csubsection}[1]{
\vspace{-0.25 cm}
\begin{center}
\stepcounter{subsection}
{\it\arabic{section}.\arabic{subsection}. #1}
\end{center}
\vspace{-0.25 cm}
}

\newcommand{\scsubsection}[1]{
\vspace{-0.25 cm}
\begin{center}
\stepcounter{subsection}
{\it #1}
\end{center}
\vspace{-0.25 cm}
}

\def\beq{\begin{equation}}
\def\eeq{\end{equation}}
\def\beqr{\begin{eqnarray}}
\def\eeqr{\end{eqnarray}}
\def\beqrs{\begin{eqnarray*}}
\def\eeqrs{\end{eqnarray*}}
\def\bet{\begin{theorem}}
\def\eet{\end{theorem}}
\def\bel{\begin{lemma}}
\def\eel{\end{lemma}}
\def\bg{\begin{figure}[tbph]\begin{center}}
\def\eg{\end{center}\end{figure}}

\def\bc{\begin{center}}
\def\ec{\end{center}}
\usepackage{amsmath}
\usepackage{accents}
\usepackage{statex}
\usepackage{multirow}
\makeatletter
\def\widebar{\accentset{{\cc@style\underline{\mskip10mu}}}}
\def\Widebar{\accentset{{\cc@style\underline{\mskip8mu}}}}
\makeatother

\def\1{\mbox{\boldmath $1$}}

\def\mS{\mathbb{S}}

\def\argmin{\mbox{argmin}}

\numberwithin{equation}{section}

{\newcommand{\killproofname}{\unskip\nopunct}}
{\newcommand{\killproofname}[1]{\unskip\aftergroup\ignorespaces\ignorespaces}}
\makeatother

\begin{document}

	\begin{center}
		{\bf\Large Estimating Extreme Value Index by Subsampling for Massive Datasets with Heavy-Tailed Distributions}\\
		
		\bigskip
		Yongxin Li$^1$, Liujun Chen$^2$, Deyuan Li$^3$ and Hansheng Wang$^4$
		
		{\it\small
			$^1$ Center for Statistical Science,  Peking University, Beijing, 100871, China;
			$^2$ School of Management, University of Science and Technology of China, Hefei, 230026, China;
			$^3$ School of Management, Fudan University, Shanghai, 200433, China; $^4$ Guanghua School of Management, Peking University, Beijing, 100871, China \\ 
			Email: yongxinli@pku.edu.cn, ljchen22@ustc.edu.cn,
			deyuanli@fudan.edu.cn,
			hansheng@gsm.pku.edu.cn
		}
	\end{center}

	\begin{singlespace}
		\begin{abstract}

			Modern statistical analyses often encounter datasets with massive sizes and heavy-tailed distributions. For datasets with massive sizes, traditional estimation methods can hardly be used to estimate the extreme value index directly. To address the issue, we propose here a subsampling-based method. Specifically, multiple subsamples are drawn from the whole dataset by using the technique of simple random subsampling with replacement. Based on each subsample, an approximate maximum likelihood estimator can be computed. The resulting estimators are then averaged to form a more accurate one. Under appropriate regularity conditions, we show theoretically that the proposed estimator is consistent and asymptotically normal.  With the help of the estimated extreme value index,  we can estimate  high-level  quantiles and tail probabilities of a  heavy-tailed  random variable consistently. Extensive simulation experiments are provided to demonstrate the promising performance of our method. A real data analysis is also presented for illustration purpose.
		\end{abstract}

		\noindent{\bf KEY WORDS:} Extreme value index; Heavy-tailed distribution; High-level quantile estimation; Massive dataset; Subsampling.

	\end{singlespace}
	
	\newpage
	
	\csection{INTRODUCTION}

	Extreme value theory (EVT) is an elegant probability theory for describing the asymptotic behavior of sample extremes (e.g., maximum or minimum). It has profound applications in many research fields. Those fields include but are not limit to risk management in geoscience \citep{katz2002statistics, an2005comparison}, finance \citep{gilliaapplication} and  public health \citep{thomas2016applications}. We refer to \cite{coles2001introduction} and \cite{beirlant2004statistics} for an excellent summary. 
	
	One problem of major interests in extreme value theory is to infer rare extreme events. 
	An example is the estimation of the high-level quantiles for the unknown population distribution. In this paper, we focus on  high-level quantile estimation for univariate heavy-tailed distributions. Mathematically, we refer to the distribution function with polynomially decaying tail as heavy-tailed \citep{hall1982some, wang2009tail, sun2019adaptive}.
	{\color{black} Specifically, we assume that for a random variable $X$ following a heavy-tailed distribution $F$, there exist two positive constants $\beta$ and $\gamma$ such that
		\begin{equation}
		\label{eq:def}
		\frac{1-F(x)}{ \beta x^{-1/\gamma}} \rightarrow 1 
		\end{equation}
		as $x\rightarrow \infty$.
		By the EVT, we know that the sample maximum in this case should follow a Fr\'echet type extreme value distribution asymptotically.
		A list of commonly used heavy-tailed distributions satisfy 	(\ref{eq:def}), including Student's t, Pareto and Fr\'echet distributions; see Section 3.1 for their specific distribution functions.} The shape of the tail distribution is mainly determined by an extreme value index $\gamma>0$  \citep{de2006extreme}. Once the parameter $\gamma$ (together with $\beta$ and some other necessary parameters) is consistently estimated, we can obtain the consistent estimations for the high-level quantiles and tail probabilities additionally.
	
	Various estimators for the extreme value index $\gamma$ have been proposed. {\color{black} Some widely-used examples include}  the Hill estimator \citep{hill1975simple} for  $\gamma>0$, the maximum likelihood estimator \citep{smith1985maximum,smith1987estimating} for $\gamma>-1/2$, the probability weighted moment method \citep{hosking1985estimation} for $\gamma<1/2$ and the moment method \citep{dekkers1989moment} for general $\gamma \in \mathbb{R}$. All those pioneer estimation methods have been found useful in the traditional setup, where the dataset is of moderate size.  Nevertheless, the story changes in the context of massive datasets. Here massive datasets refer to the type of datasets, which are too large to be read into a typical computer's memory as a whole. For instance, the default size of RAM pre-installed in a standard MacBook Pro is up to 16 GB. Hence the datasets of sizes larger than (or even close to) 16 GB can hardly be loaded into the computer's memory as a whole. Thus, they can be regarded as massive for this particular computer.  
	Massive datasets are becoming increasingly available due to the speedy advances of information technology. {\color{black} For datasets of massive sizes, they are often high-dimensional. In this case, the extreme value index needs to be estimated separately for each individual feature.}
	A direct solution is to load the whole dataset in a part-by-part  manner. Then estimators on each part are computed separately and finally aggregated to form a final estimator \citep{seila1982batching,tafazzoli2010skart,alexopoulos2019sequest}. {\color{black} However, this solution may lead to high time cost since we have to process all the data points. See the time cost for analyzing a real  massive dataset as reported in Figure \ref{workflow} for example. }
	
	To circumvent this issue,  one option is to implement a distributed parallel computing system to analyze the whole dataset in a divide-and-conquer manner \citep{lin2011aggregated, chen2014split, jordan2019communication}. However, the complexity and luxury of a distributed system  makes this  approach not suitable for all scenarios.
	Another convenient option is to draw  subsamples from the whole dataset to adapt to limited computational resources and time budgets \citep{fithian2014local,ma2015statistical,wang2018optimal, wang2021optimal}. Although subsampling may suffer from statistical efficiency reduction since it only involves a small part of data information, it is practically very attractive due to the significant benefits in terms of time savings. The savings can then
	be transferred to statistical gains in other aspects. For example, we can experiment with a broader range of candidate models in limited time budgets and update model promptly to accommodate ever-changing conditions.

	In this paper, we develop a subsample-based
	methodology to estimate the extreme value index. The general idea is to draw random subsamples with replacement from the entire dataset such that the subsamples can be loaded into the memory as a whole. To obtain a more accurate estimation,  subsampling can be performed for multiple times. With the help of subsamples, the classical peaks over threshold (POT) method \citep{davison1990models} is carefully studied. 
	Accordingly, an approximate maximum likelihood estimator for the extreme value index can be computed for each subsample. Next, estimators obtained from different subsamples are averaged together so that a more powerful estimator can be constructed. We refer to the final estimator as the averaged maximum likelihood (AML) estimator. The consistency and the asymptotic normality of our new estimator are also established. Based on the AML estimator, we are 
	able to estimate high-level quantiles for the underlying data distribution consistently.

	It is noteworthy that although multiple subsamples are involved in our approach, the number of subsamples can be  relatively small if the execution efficiency is the major concern (which is often the case in real applications). Therefore, the overall amount of data points included in all subsamples is still much smaller compared to the whole dataset. This leads to a significant reduction in computation and data loading overhead. In fact, as we will elaborate in Sections 2 and 3, the number of
	subsamples is allowed to be varied within a wide range. Its value can be determined according to several factors including the whole dataset size, the memory capacity, expected statistical accuracy and time budgets. Consequently, the tuning parameter selection in our method is more practical and flexible for massive data analysis.

	The rest of this paper is organized as follows. Section 2 introduces an averaged maximum likelihood method and investigates its asymptotic properties. Numerical studies  are given in Section 3. A real airline dataset analysis is included in Section 4. Finally, the paper is concluded with a brief discussion in Section 5. All the proofs, together with additional lemmas, are deferred to the Appendix.

	\csection{THE METHODOLOGY}
	\csubsection{Averaged Maximum Likelihood Estimation}
	
	Let $X_i$'s $\in \mathbb{R} \ (i=1,2,\cdots,N)$ be independent and identically distributed copies of a random variable $X \in \mathbb{R} $ with cumulative distribution function $F$.
	In this work, we focus on heavy-tailed distributions only. We  follow the tradition \citep{hall1982some,wang2009tail,sun2019adaptive} and assume that the decaying rate of the distribution's tail probability is polynomial. Specifically, we assume that $F$ satisfies (\ref{eq:def}). This suggests that high-level quantiles estimates can be obtained  as long as the parameter $\gamma$ (or plus $\beta$) can be consistently estimated.

	To estimate the extreme value index $\gamma$, a classical method is to model the conditional excess probability over a high threshold. This technique is called the peaks over threshold (POT) method  \citep{davison1990models}. Specifically, for a random variable $X$ satisfying (\ref{eq:def}) and a given high threshold $u$,  it can be shown that 
	\begin{equation}
	\label{eq:quantile}
	\lim_{u\rightarrow \infty}P(X>tu|X>u) = \lim_{u\rightarrow \infty}\frac{1-F(tu)}{1-F(u)}  = t^{-1/\gamma}, 
	\end{equation}
	for $t>1$. In other words, the exceedances over the high threshold $u$ can be approximated by a Generalized Pareto distribution with shape parameter $\gamma$ \citep{pickands1975statistical,coles2001introduction}. 
	
	Inspired by the classical POT method, we develop here a subsample-based counterpart. To this end,  denote the index set of the full dataset as $\mS= \left\{1,\cdots,N\right\}$. We randomly draw $K$ subsamples of size $n$ with replacement from the full dataset. For the $k$-th subsample ($1\leq k \leq K$), the correspnding index set is denoted as $\mathcal{S}_{k} = \{m^{(k)}_1,m^{(k)}_2,\cdots , m^{(k)}_n\} \subset \mS$. Conditioning on $\mS$, $\{m^{(k)}_i: 1\leq i \leq n, 1 \leq k \leq K\}$
	are independent and identically distributed with $P(m^{(k)}_i=j)=1/N$ for any  $j \in \mS$. 
	Let $f(x|u)$ be the probability density function of $X$ given $X>u$. {\color{black} By (\ref{eq:quantile}), $f(x|u) = (\gamma u)^{-1} (x/u)^{-1/\gamma-1} + o(1)$ for $u\rightarrow \infty$ and $x>u$}. Then we can derive the following approximate log-likelihood function for the $k$-the subsample as
	\begin{eqnarray*}
		\label{eq:amle}
		\mathcal{T}_k(\gamma)= \sum_{i \in \mathcal{S}_k} \Big\{-\log \gamma - \log u - \big(1+ 1/\gamma\big) \log\big(X_{i}/u\big)  \Big\}I\big(X_{i}>u\big).
	\end{eqnarray*}
	By maximizing $\mathcal{T}_k(\gamma)$ with respect to $\gamma$, we get an approximate maximum likelihood estimator as $\hat{\gamma}_{k} = \arg\max\limits_{\gamma}  \mathcal{T}_k(\gamma)=(n_{k}^u)^{-1} \sum_{i \in \mathcal{S}_k} \log(X_{i}/u)I(X_{i}>u)$. Here, $n_{k}^u = \sum_{i \in \mathcal{S}_k} I(X_{i}>u)$ is  the number of observations over a pre-defined threshold $u$ for the $k$-th ($1\leq k \leq K$) subsample. We then combine these subsample-based estimators together. This leads to the averaged final estimator 
	\begin{eqnarray*}
		\hat{\gamma}_{\text{\sc aml}}=K^{-1}\sum_{k=1}^K \hat{\gamma}_{k}=K^{-1}\sum_{k=1}^K (n_{k}^u)^{-1} \sum_{{i \in \mathcal{S}_k}} \log\big(X_{i}/u\big)I\big(X_{i}>u\big).
	\end{eqnarray*}
	We refer to $\hat{\gamma}_{\text{\sc aml}}$ as the averaged maximum likelihood (AML) estimator. As we can see, the AML estimator is closely related to the classical Hill estimator \citep{hill1975simple} in this setting.  A more general combination scheme is to use the weighted average of $\hat{\gamma}_{k}$s. For example, the weight for subsample $\mathcal{S}_k$ can be selected to be proportional to $n_k^u$ or its transformation. This leads to much improved finite sample performance if unequal subsample sizes are allowed. See Section 3.2 for numerical evidences.	
	
	\csubsection{Theoretical Properties}
	\label{sec:theoretical_properties}
	
	To investigate the asymptotic behavior of $\hat{\gamma}_{k}$ and the averaged estimator $\hat{\gamma}_{\text{\sc aml}}$, the following conditions are needed.
	\begin{itemize}
		\item [(C1)] {(\sc Tail Probability)}  {\color{black} Define $\alpha_u=P(X>u)$.  Assume that as $u \rightarrow \infty$, there exists  a set of fixed constants $\beta>0, \gamma>0, \delta>0$ and $C \in \mathbb{R}$ such that $\alpha_u= \beta u^{-1/\gamma}\left\{1+Cu^{-\delta}+\\ o(u^{-\delta})\right\}$.}
		
		\item [(C2)] {(\sc Subsample Ratio)} As $N \rightarrow \infty$, assume that $n\rightarrow \infty$ and $n=o(N)$.
		
		\item [(C3)] {(\sc Divergence Rate)}  As $N\rightarrow \infty$, we assume that the threshold $u$ satisfies $u\rightarrow \infty$, $un^{-\gamma}\rightarrow 0$ and $un^{-\gamma/(2\delta\gamma + 1)} \rightarrow \infty$.
		
		\item [(C4)] {(\sc Number of Subsamples)} As $N \rightarrow \infty$, we assume that  $K=o(\min\{n^{-1}u^{1/\gamma+2\delta}, nu^{-1/\gamma}\})$.
		
	\end{itemize}
	
	Condition (C1) quantifies the decaying speed of the tail probability. A similar condition has been popularly assumed in extreme value literature \citep{hall1982some, smith1987estimating,resnick2007heavy,wang2009tail}. It is equivalent to replace $\beta$ and $C$ in Condition (C1) by two smooth functions $\beta(u)$ and $C(u)$ respectively, as long as $\beta(u)\rightarrow \beta$ and $C(u)\rightarrow C$ as $u\rightarrow \infty$. 
	Condition (C2) is  widely used in subsampling literature \citep{kleiner2014scalable,wang2018optimal, wang2020optimal}. It enforces that the subsample size $n$ should diverge to infinity at an appropriate speed. Specifically, $n$ should not be too small. Otherwise, the asymptotic theory cannot be developed. On the other hand, $n$ should not be too large. Otherwise the subsample is  too large to be loaded into memory. Condition (C3) imposes a restriction on the divergence speed of the threshold level $u$. On the one side, it should not be too small. Otherwise the tail probability cannot be well approximated by the polynomial function as given in Condition (C1). On the other side, it cannot be too large either. Otherwise 
	the exceedance size of each subsample $n_k^u$
	is too small to support a consistent estimator.
	Condition (C4) implies that the number of subsamples $K$ can be either a fixed integer or diverge to infinity but not too fast. Obviously, larger $K$ leads to smaller variability of  $\hat{\gamma}_{\text{\sc aml}}$. Since 
	the local estimators $\hat{\gamma}_k$s are identically distributed,  $\hat{\gamma}_{\text{\sc aml}}$ shares the same amount of bias as $\hat{\gamma}_{k}$. Accordingly, the bias of  $\hat{\gamma}_{\text{\sc aml}}$ cannot be reduced by increasing $K$. As a consequence, if $K$ is too large, the bias suffered by $\hat{\gamma}_{\text{\sc aml}}$ becomes nonnegligible, as compared with its standard deviation. Consequently, 
	the number of subsamples $K$ cannot be too large either. Otherwise the benefit introduced by larger $K$ in terms of variability reduction can be completely offset by its bias. 

	\begin{theorem}
		\label{th:sub}
		Let $n_{*}^u = \sum_{k=1}^K n_k^u$ be the size of the total exceedances. Assume Conditions (C1)--(C4) hold, then we have (1) $\sqrt{n_k^u}(\hat{\gamma}_{k}-\gamma) \xrightarrow{d} N(0,\gamma^2)$ for $1\leq k \leq K$;
		(2) $\sqrt{n_{*}^u}(\hat{\gamma}_{\text{\sc aml}}-\gamma) \xrightarrow{d} N(0,\gamma^2)$.	
	\end{theorem}
	
	According to Theorem \ref{th:sub}, we know that both the local estimator $\hat{\gamma}_k$ and the averaged estimator $\hat{\gamma}_{\text{\sc aml}}$ are consistent and asymptotically normal. The rate of convergence is $\sqrt{n_k^u}$ for $\hat{\gamma}_k$  and $\sqrt{n_{*}^u}$ for  $\hat{\gamma}_{\text{\sc aml}}$, respectively. To make inference about $\gamma$, consider a confidence level $1-\alpha$. An asymptotically valid confidence interval for $\gamma$ can be constructed as  $\hat{\gamma}_{\text{\sc aml}} \pm \Phi^{-1}(1-\alpha/2)\hat{\gamma}_{\text{\sc aml}}/\sqrt{n_*^u}$, where $\Phi$ is the cumulative distribution function of the standard normal distribution. In the meanwhile, assume a high threshold $u>0$, (\ref{eq:quantile}) indicates that $ P(X>x) \approx \alpha_u \left(x/u\right)^{-1/\gamma}
	$ for any $x>u$. Therefore, for a quantile level $1-\tau$ with sufficiently small positive $\tau$, we have
	$P(X > q^{(a)}_{1-\tau}) \approx 
	\tau$ with $q^{(a)}_{1-\tau} = u\left(\alpha_u/\tau\right)^{\gamma}$. Denote the exact $(1-\tau)$-th quantile of distribution $F$ as $q_{1-\tau}$. That is $P(X > q_{1-\tau})=\tau$. We should expect $q^{(a)}_{1-\tau} \approx q_{1-\tau}$.
	A natural estimator for $q^{(a)}_{1-\tau} $ is given by $\hat{q}^{(a)}_{1-\tau} = u\left(\hat{\alpha}_u/\tau\right)^{\hat{\gamma}_{\text{\sc aml}}}$ with $\hat{\alpha}_u = (nK)^{-1}n_*^u$, provided $\tau < \alpha_u$. This leads to a useful method to approximate the high-level quantile for $X$, if $\gamma$ can be consistently estimated. 
	Let $\hat{\tau} = P(X > \hat{q}^{(a)}_{1-\tau})$, The following theorem describes the asymptotic behavior of  $\hat{q}^{(a)}_{1-\tau}$ and $\hat{\tau}$.

	\begin{theorem}
		\label{th2}
		Assume  $0<\tau <\alpha_u$, $\log(\alpha_u/\tau)=o(\sqrt{nK\alpha_u})$, and Conditions (C1)--(C4) hold, then we have (i) $\hat{q}^{(a)}_{1-\tau}/{q}_{1-\tau} \xrightarrow{p} 1$; and (ii)  $\hat{\tau}/\tau \xrightarrow{p}1$.
	\end{theorem}
	
	Theorem \ref{th2} indicates that we can estimate a very high-level quantile $q_{1-\tau}$ consistently. The resulting tail probability $\hat{\tau}$ is also ratio-consistent for the intended tail probability. It is remarkable that this can be done even if $\tau$ is too small to have sufficient number of extreme observations (i.e. $X_i$s such that $X_i>q_{1-\tau}$).
	Take an example for illustration. {\color{black} Consider a low level} $\tau = 10^{-5}$. To estimate the corresponding quantile $q_{1-\tau}$, very large sample size (e.g. $N\gg10^5$) is needed if a traditional nonparametric method is used. However, under the help of  $\hat{\gamma}_{\text{\sc aml}}$, we can estimate $q_{1-\tau}$ consistently with $K=10$ and $n=10^3$. Accordingly, the tail probability $\tau$ can also be estimated consistently. 
	
	\csection{NUMERICAL STUDIES}
	
	\csubsection{Simulation Models}
	\label{sec:simulation_models}
	
	To demonstrate the finite sample performance of the proposed estimators, we conduct a number of simulation studies in this section. To this end, we need to generate $X_i$s from a distribution whose tail probability satisfies Condition (C1).
	Specifically, we consider the following examples.

	{\sc Example 1.} {({\sc Student's} {\it t}{\sc-distribution)}}
	The first example considered here is the student's {\it t}-distribution. It has been popularly used in extreme value modeling literature \citep{blattberg1974comparison,stoyanov2011fat}.
	The distribution is heavy-tailed on both sides (right and left). Specifically, let $X$ follow a Student's {\it t}-distribution with $v$ degrees of freedom,  then the right tail probability becomes \citep{beirlant2004statistics} 
	\begin{eqnarray*}
		P\big(X>u\big) & = & \int_{u}^{\infty} \frac{\Gamma\left(\frac{v+1}{2}\right)}{\sqrt{v \pi} \Gamma\left(\frac{v}{2}\right)}\left(1+\frac{x^{2}}{v}\right)^{-\frac{v+1}{2}} dx \\
		& = & \frac{\Gamma\left(\frac{v+1}{2}\right)v^{(v-1)/2}}{\sqrt{v \pi} \Gamma\left(\frac{v}{2}\right)}
		u^{-v}\left\{1-\frac{v^2(v+1)}{2(v+2)}u^{-2}+o(u^{-2})\right\}.
	\end{eqnarray*}
	Consequently, the corresponding extreme value index is given by $\gamma = 1/v$.
	The degree of freedom parameter $v$ controls 
	the tail behavior of  the Student's {\it t}-distribution. In this simulation study, we consider Student's {\it t}-distributions with $v=1$ and $v=2$, denoted as $t(1)$ and $t(2)$.

	{\sc Example 2.} {(\sc Pareto distribution}) Pareto distribution is another widely adopted heavy-tailed distribution. It is found to be particularly useful in income distribution research  \citep{wold1957model,nirei2016pareto}. The right tail probability of  $\text{Pareto}(x_m,\alpha)$ where $\alpha>0$ is given by $P(X>u)=(u/x_m)^{-\alpha}$ with $u \geq x_m$ .
	Accordingly, the extreme value index is given by $\gamma = 1/\alpha$. Specially, the parameter $C$ in Condition (C1) is equal to 0. 
	The tail heaviness is controlled by the parameter $\alpha$. In this simulation study, we consider cases $\text{Pareto}(2,1)$ and $\text{Pareto}(2,2)$.
	
	{\sc Example 3.} {(\sc Fr\'echet distribution}) We study here the Fr\'echet distribution. The tail probability of  $\text{Fr\'echet}(\alpha)$ where $\alpha>0$ is given by $P(X>u)=1-\exp(-u^{-\alpha})=u^{-\alpha}\left\{1-0.5u^{-\alpha}+o(u^{-\alpha})\right\}$. As a consequence, the extreme value index is given by $\gamma = 1/\alpha$. Thus, the tail behavior of a  Fr\'echet distribution is determined by $\alpha$. In this simulation study, we consider Fr\'echet(1) and Fr\'echet(2) cases.

\begin{figure}[!ht]
		\centering
		\includegraphics[scale=.8]{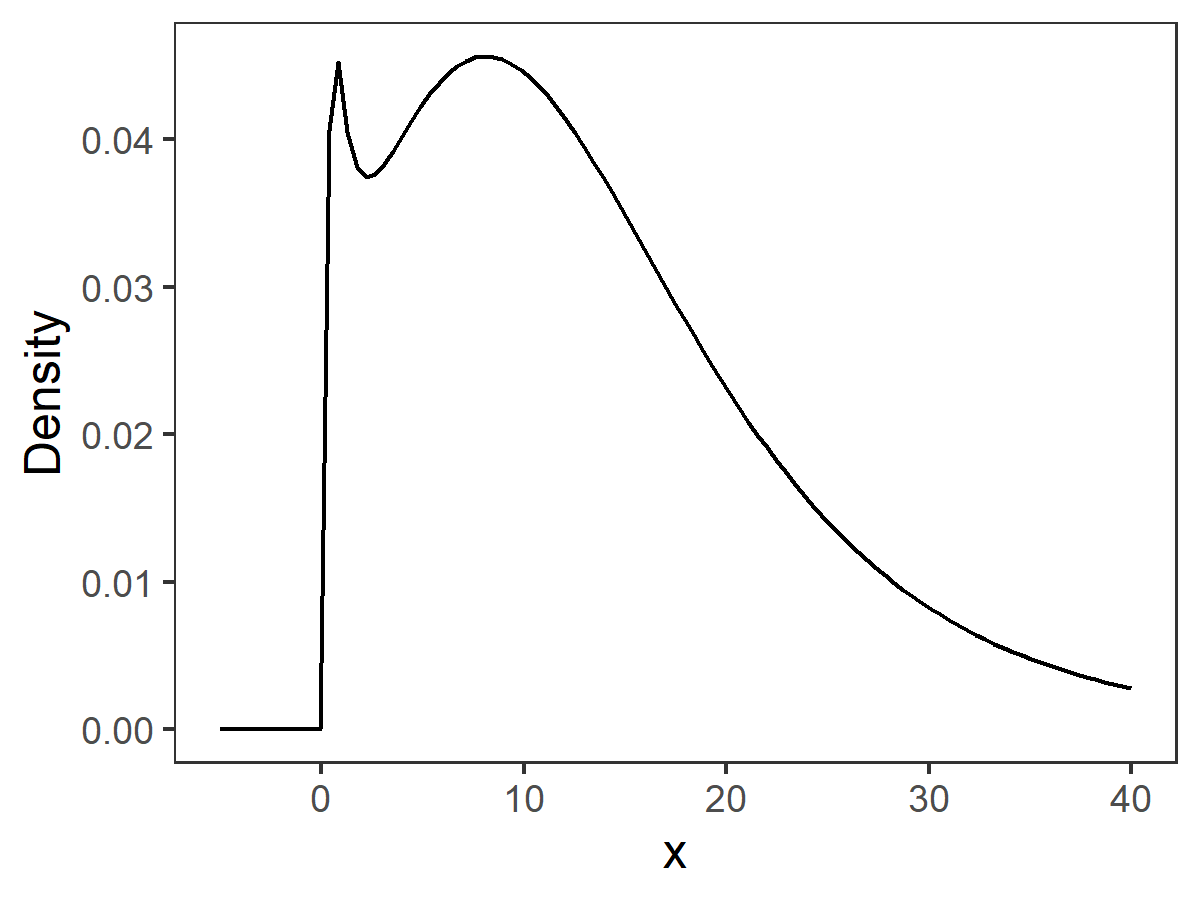}
		\caption{Density plot  of the multimodal generating distribution in Example 4.} 
		\label{multimodal} 
	\end{figure}

	{\sc Example 4.} {(\sc Multimodal distribution)} We consider in this example a multimodal distribution.
	Inspired by \cite{Cao2020evt}, we adopt the following method to generate $X_i$s from a heavy-tailed multimodal distribution. Specifically, $\{X_i\}_{i=1}^N$ is independently generated by $X_i = \max\{Y_i,Z_i\}$, where $Y_i$ and $Z_i$ are independently generated from Fr\'echet(1) and Gumbel(8,8), respectively. Here, $\text{Gumbel}(\alpha,\beta)$ refers to the Gumbel distribution with tail probability $P(Z>u) =1 - \exp\{-\exp{(Z-\alpha)/\beta}\}$. It is obvious that the tail probability of the Gumbel distribution diverges to zero exponentially. As a consequence, it can be shown that  $X_i$s satisfies Condition (C1). The corresponding extreme value index is determined by the Fr\'echet part and thus equals to 1. The density function of $X_i$ is plotted in Figure \ref{multimodal}, which is multimodal.
	
	{\sc Example 5.} {(\sc Non-independent data}) We consider here a non-independent data example. Assume that  $\{X_i\}_{i=1}^{N}$ are partitioned into $m$ clusters.
	The $k$-th cluster contains $N_k$ random variables, which is denoted by $\{X_{k1},\cdots,X_{k N_k}\}, 1\leq k \leq m$.  For cluster $k$, $X_{kj}$ is generated by $X_{kj} =  \alpha_k + \varepsilon_{kj}, 1\leq j \leq N_k$. This is a standard random effects model,  where the random effect parameter $\alpha_k$ is generated from Fr\'echet(1) and the random noise $\varepsilon_{kj}$ is generated from $N(0,1)$.  Consequently, $X_{kj}$s within the same cluster are correlated.  It can be verified that the tail behavior of $X_{kj}$ is mainly decided by
	$\alpha_k$. Thus, the corresponding extreme value index is 1.

	The above examples specify the generating distribution for $X$. Once the generating distribution and the whole sample size $N$ are given, we can simulate the full dataset. Here, we consider different $N$ values for different examples. For Examples 1--3, we set $N=10^5, 5\times 10^5, 10^6, 5\times10^6$. While for Examples 4--5, we set $N=10^7, 4\times 10^7, 7\times 10^7, 10^8$. In view of Condition (C2) requiring that $n=o(N)$, we set $n=\lfloor N^{0.5} \rfloor$. Here, $\lfloor x \rfloor$ stands for the largest integer no larger than $x$. Condition (C3) requires that $u=n^{\gamma/(1+h\gamma )}$ with $h\in(0,2\delta)$. It implies that $P(X>u)=O(n^{-1/(1+h\gamma)})$. Therefore, the threshold value is set to be the $(1-n^{-1/(1+h\gamma)})$-th quantile of the given generating distribution. 
	We fix $h=0.8\delta$ for Examples 1--3 and $h=0.6 \delta$ for Examples 4--5.
	Moreover, Condition (C4) indicates that the divergence speed of $K$ is at most $o(n^{\delta/(\gamma^{-1}+\delta)})$. Hence, we let $K=  \lfloor n^{C_K\delta/(\gamma^{-1}+\delta)}\rfloor $ with coefficients $C_K$=0.3, 0.5, 0.7 (for Examples 1--3) or $C_K$=0.3, 0.4, 0.5 (for Examples 4--5). The parameter $\delta$ of Examples 1--5 is set to be 2, 5, $\alpha$, 1 and 1 respectively. Thereafter, the averaged estimator $\hat{\gamma}_{\text{\sc aml}}$ can be computed.

	\csubsection{Simulation Results}
	
	To obtain a reliable evaluation, 
	we randomly replicate each simulation experiment for a total of $R=1,000$ times. 
	Let $\hat{\gamma}_{\text{\sc aml}}^{(r)}$ be the averaged estimator obtained in the $r$-th ($1\leq r \leq R$) replication. We then define the root mean squared error (RMSE) as
	\begin{eqnarray*}
		\text{RMSE}(\hat{\gamma}_{\text{\sc aml}})=\left\{R^{-1}\sum_{r=1}^R\Big(\hat{\gamma}^{(r)}_{\text{\sc aml}}-\gamma\Big)^2\right\}^{1/2}.
	\end{eqnarray*}
	Additionally, an asymptotic $1-\alpha$ confidence interval for  $\gamma$ can be constructed as $\text{CI}^{(r)}=\Big(\hat{\gamma}_{\text{\sc aml}}^{(r)}-\Phi^{-1}(1-\alpha/2)\hat{\gamma}_{\text{\sc aml}}^{(r)}/\sqrt{n_{*(r)}^{u}}, \hat{\gamma}_{\text{\sc aml}}^{(r)}+\Phi^{-1}(1-\alpha/2)\hat{\gamma}_{\text{\sc aml}}^{(r)}/\sqrt{n_{*(r)}^u}\Big)$. Then the empirical  coverage probability is given by $\text{ECP}=R^{-1}\sum_{r=1}^R I(\gamma \in \text{CI}^{(r)})$. Here we fix $\alpha=0.05$.
	We can further obtain a $(1-\tau)$-th quantile  estimator for $X$ in the $r$-th replication as $\hat{q}^{(a,r)}_{1-\tau}$. We can compute the tail probability $\hat{\tau}^{(r)}=P\big(X>\hat{q}^{(a,r)}_{1-\tau}\big)$ according to its theoretical formula. This leads to relative accuracy (RA) for tail probability as
	\begin{eqnarray*}
		\text{RA}=\left\{R^{-1}\sum_{r=1}^R \Big(\hat{\tau}^{(r)}/\tau -1\Big)^2\right\}^{1/2}.
	\end{eqnarray*}
	In this simulation study, we set $\tau=10^{-3}$.
	
	The simulation results are summarized in Tables 1--4 (in Appendix B). First,  we find that the RMSE of $\hat{\gamma}_{\text{\sc aml}}$ decreases towards 0 as
	the whole exceedances size $n_*^u$ increases for all the examples. In particular, the standard deviation  decreases towards 0 as $n_*^u$ goes to infinity. 
	However, with a fixed whole sample size $N$, the bias of $\hat{\gamma}_{\text{\sc aml}}$ has little changes as the number of subsamples $K$ increases. This is expected  because the bias is mainly controlled by $n_*^u$ and has nothing to do with $K$.
	Second,  the empirical coverage probabilities (ECP)  are all quite close to the nominal level 95\%. This confirms the asymptotical normality of $\hat{\gamma}_{\text{\sc aml}}$. Additionally, the relative accuracy of the tail probability estimator  (the last column) approaches 0 as $n_*^u$ increases towards infinity. This corroborates the theoretical findings given in Theorem \ref{th2}.  Additionally, although we require that $K$ goes to infinity in Condition (C4), the simulation results shows that in practice, a relative small $K$ (corresponds to $C_K =0.3$) is large enough for a practically acceptable statistical performance. Hence, if the statistical consistency of the estimator and execution efficiency is the major concern (which is indeed often the case in practice), then a relatively small $K$ can be used.

 We next evaluate the performance of two different combination schemes in terms of  $\hat{\gamma}_{k}$s. Apart from the simple average based estimator as we introduced in Section 2.1, we also consider to use a weighted average of  $\hat{\gamma}_{k}$s to form the final estimator. Specifically, the weight for the $k$-th subsample is proportional to its exceedance size $n_k^u$. 
		We select $t(1)$ as the generating function and follow a similar model setup as stated in Section 3.1. Specifically, we set $N=10^6, 5\times 10^6, 10^7$ and $n = \lfloor N^{0.5} \rfloor$.
		The number of subsamples $K$ is set to be  $K=10,14,20$ respectively .
		The threshold value $u$ satisfies $P(X>u) = O(n^{-1/(1+2h)})$ with $h=0.8$. We consider two different scenarios  regarding the subsample size $n_k$. The first scenario is a balanced case with $n_k=n$ for each $k$. The second scenario is an imbalanced case with $n_k=1.5n$ for $1\leq k \leq K/2$ and $n_k=0.5n$ for $1+K/2 \leq k \leq K$. We replicate the simulation for 1,000 times and report the  results in Table \ref{tabX}. It can be seen that  estimators produced by the simple average and weighted average schemes  perform  similarly, if equal subsample sizes are used. 
		In contrast, the weighted average estimator performs much better than the simple average one,  if very different subsample sizes are allowed.

	\csubsection{The Competing Estimators}
	
	For comparison purpose, we also evaluate here a number of competing estimators. They are, respectively, the probability weighted moment estimator \citep{hosking1987parameter} and  the  moment  estimator \citep{dekkers1989moment}. {\color{black} These two methods are constructed according to the method of moments. In contrast, our estimator is developed based on the approximate likelihood (\ref{eq:amle}) and is more closely related to the Hill estimator \citep{hill1975simple}}. Thus, the classical Hill estimator is not included for comparison. 
	Similar to the AML estimator, the moment estimator and the probability weighted moment estimator of $\gamma$ can be computed on each subsample. This leads to a total of $K$ estimators. They are then averaged to form the final ones.
	The resulting estimators are referred to as an averaged  probability weighted moment (APWM) estimator and an averaged moment (AMO) estimator respectively.  Specifically, for the $k$-th subsample, let $X^{(k)}_{1:n}\leq X^{(k)}_{2:n} \leq \cdots \leq X^{(k)}_{n:n}$ denote the  order statistics.
	Then the APWM estimator
	is given by 
	\begin{equation}
	\hat{\gamma}_{\text{\sc apwm}} = K^{-1} \sum_{k=1}^K \hat{\gamma}_{\text{\sc pwm},k} = K - \sum_{k=1}^K \frac{2Q_n^{(k)}}{P_n^{(k)} - 2Q_n^{(k)}},
	\end{equation}
	where $P_n^{(k)} = (n_k^u)^{-1} \sum_{i=1}^{n_k^u} X^{(k)}_{n-i+1:n}- u$, $Q_n^{(k)} = (n_k^u)^{-1}  \sum_{i=1}^{n_k^u} \{(i-1)/(n_k^u -1)\} (X^{(k)}_{n-i+1:n}-u)$. The AMO estimator is given by 
	\begin{equation}
	\hat{\gamma}_{\text{\sc amo}} = K^{-1} \sum_{k=1}^K \hat{\gamma}_{\text{\sc mo},k}  = K +  \sum_{k=1}^K M_{n,1}^{(k)} - \frac{1}{2}\sum_{k=1}^K \left(1-\frac{\left(M_{n,1}^{(k)}\right)^2}{M_{n,2}^{(k)}}\right)^{-1},
	\end{equation}
	where $M_{n,l}^{(k)} = (n_k^u)^{-1} \sum_{i=1}^{n_k^u} (\log X^{(k)}_{n-i+1:n} - \log u)^l$, $l=1,2$.

\begin{figure}[!ht]
		\centering
		\includegraphics[scale=1]{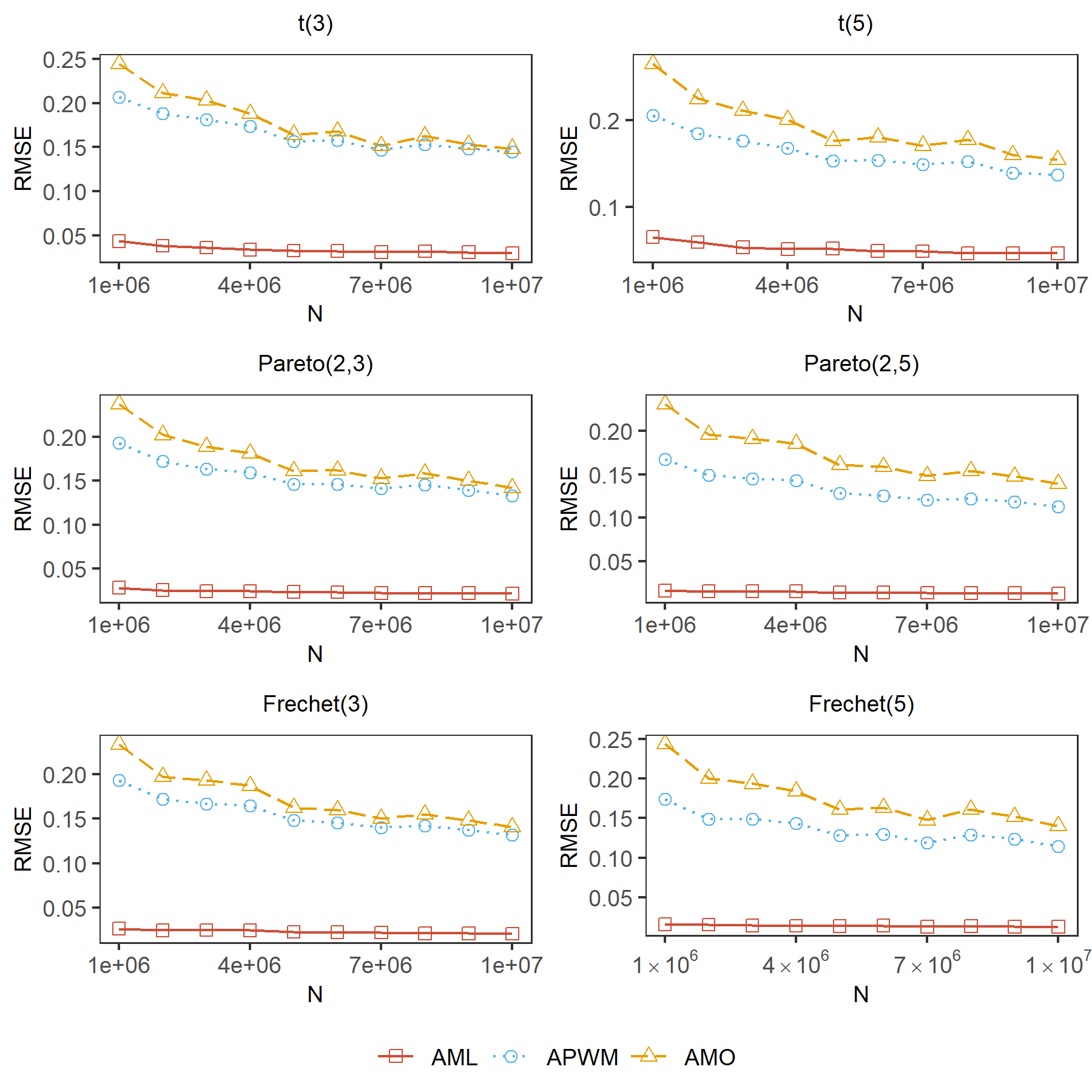}
		\caption{Root mean squared error (RMSE)
			of various estimators with $N$ ranging from $10^6$ to $10^7$, $n=\lfloor N^{0.5} \rfloor$ and $K=10$. The RMSE is computed based on 1,000 replications. Different plots correspond to different distributions.} 
		\label{fig1} 
	\end{figure}

	Note that the classical probability weighted moment estimator is consistent only for $\gamma<1$ and is asymptotically normal only for $\gamma<1/2$. For a fair comparison, we only consider here generating distributions with $\gamma < 1/2$. Specifically, those distributions are 
	t(3), t(5), Pareto(2,3), Pareto(2,5), Fr\'echet(3) and  Fr\'echet(5), respectively. The whole sample size $N$ varies form $10^6$ to $10^7$. The subsample size $n$ is fixed at $n=\lfloor N^{0.5} \rfloor$. The number of subsamples is given by  $K=10$. The threshold value $u$ is set to be the  $(1-n^{-0.6})$-th quantile of $X$. We compare the RMSE values of $\hat{\gamma}$ for three different estimation methods (AML, AMO and APWM). The simulation results are shown in Figure \ref{fig1}. From Figure \ref{fig1}, we can observe that the AML estimator always performs  best in all cases.
	
\begin{figure}[!ht]
		\includegraphics[scale=.24]{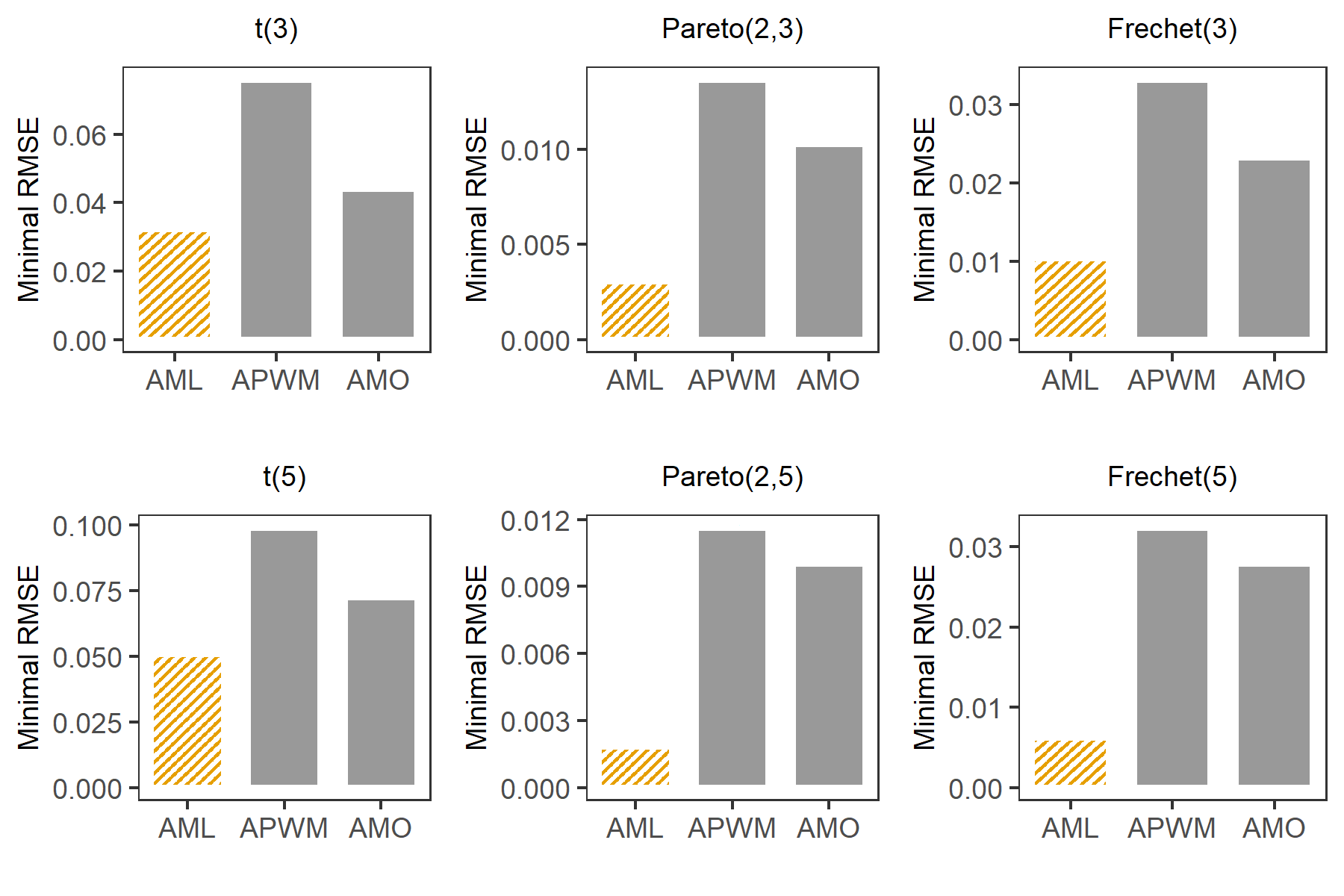}\par
		\caption{The minimal RMSE values of three estimators under different distributions. In each plot, different bars correspond to different estimators. The bar heights are equal to the minimal RMSE values, which is computed across 50 different threshold values. The bar with diagonal-stripes pattern represents the estimator with the global minimal RMSE value. In this study, $N=5\times10^6$, $n=2,236$ and $K=10$. 
			The RMSE is computed based on 1,000 replications.} 
		\label{fig2}
	\end{figure}

	However, it is remarkable that the optimal threshold value for different estimators could be different. Thus, comparing the performance of different estimators under one pre-specified threshold value might be unconvincing. To address the issue, we fix $N=5\times10^6$ and $n=\lfloor N^{0.5} \rfloor=2,236$. We compare the best finite sample performances of the three estimators. That is their minimal RMSE values across a wide range of threshold values. In this study, a total of 50 threshold values are considered. Their corresponding tail probabilities are given by $0.5 + i/100$ for $i=0,1\cdots,49$. We present the detailed results in Figure \ref{fig2}. As one can see, the minimal RMSE values of the AML estimators remain the smallest.

	\csubsection{Threshold Selection}
	\label{subsec: threshold}

	We now consider the problem  of the threshold selection for the AML estimation.  As we pointed out earlier in  Section 2.2, it is critical to specify an appropriate $u$. On the one hand, $u$ should be sufficiently large to ensure  that the error in approximating tail probability by Condition (C1) is small enough. On the other hand, larger $u$ corresponds to small exceedance size (i.e., $n_*^u$), which leads to poor statistical performances of AML estimators. In this section, we follow \cite{wang2009tail} and provide here a simple but effective framework to select the threshold $u$ adaptively. Concretely, note that given a proper $u$, the distribution of $\gamma^{-1}\log(X/u)$ conditional on $X>u$ is approximately standard exponential. This implies that the distribution of $\exp\{-\gamma^{-1}\log(X/u)\}$ conditional on $X>u$ is approximately a uniform distribution on $[0,1]$. As a result,  an optimal threshold value $u$ {\color{black} should be the one corresponding to} the minimal distance between the  uniform distribution on $[0,1]$ and the empirical distribution  of $\{\hat{Z}_i:X_i >u, i \in \cup_{k=1}^K \mathcal{S}_k\}$, where $\hat{Z}_i = \exp\{-(\hat{\gamma}_{\text{\sc aml}})^{-1}\log(X_i/u)\}$. 
	
	There are many methods to quantify the distance between the empirical distribution of the sample and the reference distribution. One popular method is to apply the Cram\'er-von Mises statistic \citep{choulakian2001goodness}. Based on a total of $K$ subsamples,  it can be calculated as follows: 
	\begin{eqnarray}
	\label{eq:cvm}
	W^2_1(u) = \sum_{i=1}^{n_*^u} \left\{{\hat{Z}_{i:n_*^u}} - (2i-1)/\big(2n_*^u\big)\right\}^2  + 1/\big(12n_*^u\big)
	\end{eqnarray}
	where $n_*^u = \sum_k n_k^u = \sum_k \sum_{i \in \mathcal{S}_k} I(X_i>u)$ denotes the size of total exceedances as defined in Theorem \ref{th:sub}. $\hat{Z}_{i:n_*^u}$ represents the $i$-th order statistic of $\{\hat{Z}_i\}$ with 
	$\hat{Z}_{1:n_*^u} \leq \hat{Z}_{2:n_*^u} \leq \cdots \leq  \hat{Z}_{n_*^u:n_*^u}$.
	Note that if 
	$\hat{Z}_{i:n_*^u}$ is indeed uniformly distributed on $[0,1]$, then the value $W^2(u)$ should be small. Hence, we suggest that the optimal threshold value $u$ can be selected as $u_1^* =  \argmin_{u}  W_1^2(u)$. Note that 
	the computation scheme of (\ref{eq:cvm}) requires going through $K$ subsamples twice. To reduce time overhead costs, a simpler scheme is to calculate $W^2$ based on the first 
	subsample $\mathcal{S}_1$, where the corresponding selected threshold value is defined as $u_2^*$. For elaboration convenience, we refer to the procedures of selecting $u_1^*$ and $u_2^*$ as \textsf{Scheme 1} and \textsf{Scheme 2} respectively.

\begin{figure}[!ht]
		\centering
		\begin{subfigure}{0.49\textwidth}
			\includegraphics[width=\textwidth]{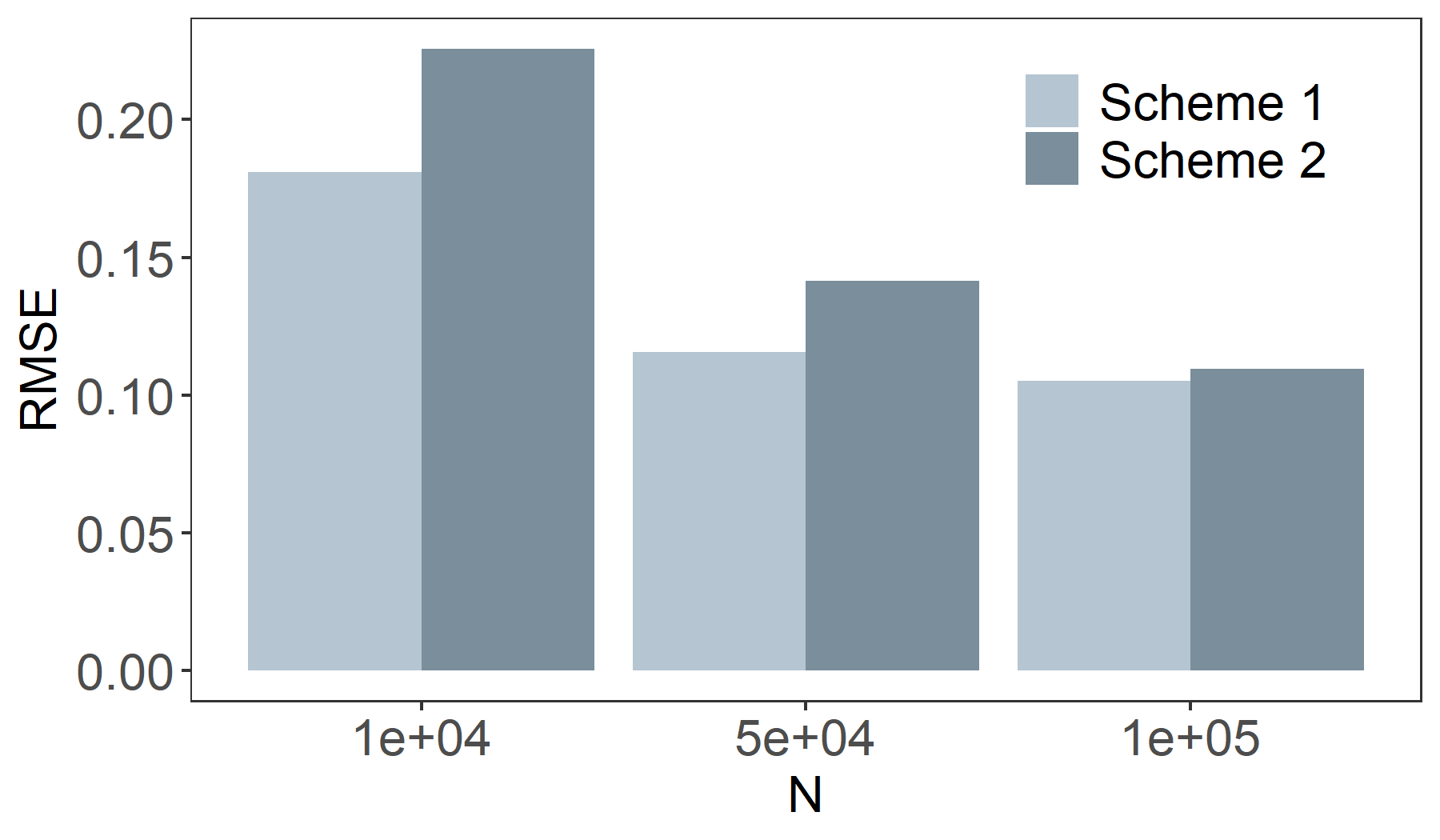}
			\caption{$t(1)$ and $K = \lfloor n^{0.4}\rfloor$.}
		\end{subfigure}
		\begin{subfigure}{0.49\textwidth}
			\includegraphics[width=\textwidth]{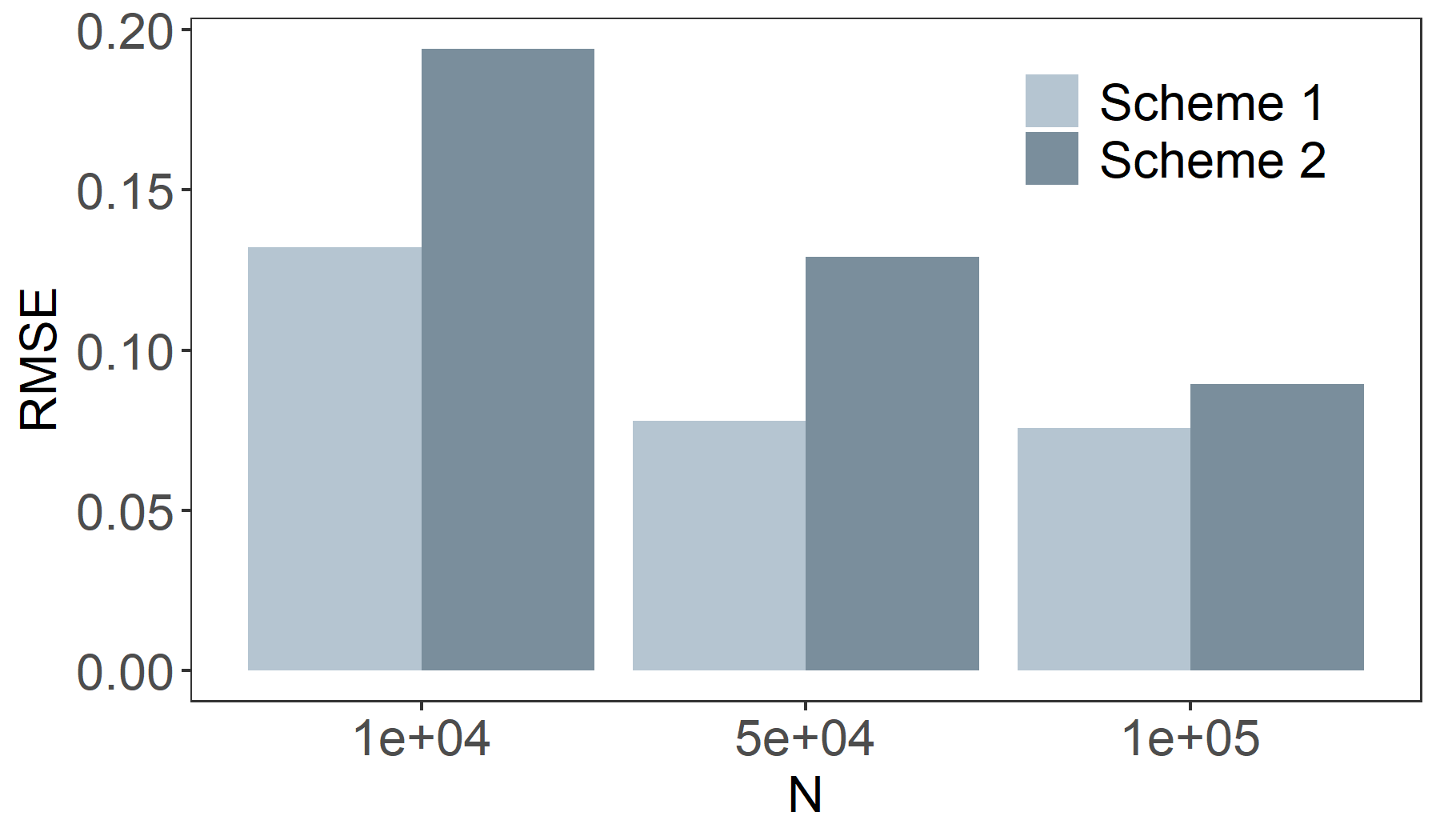}
			\caption{$t(1)$ and $K = \lfloor n^{0.6}\rfloor$. }
		\end{subfigure}  \\[3mm]
		\begin{subfigure}{0.49\textwidth}
			\includegraphics[width=\textwidth]{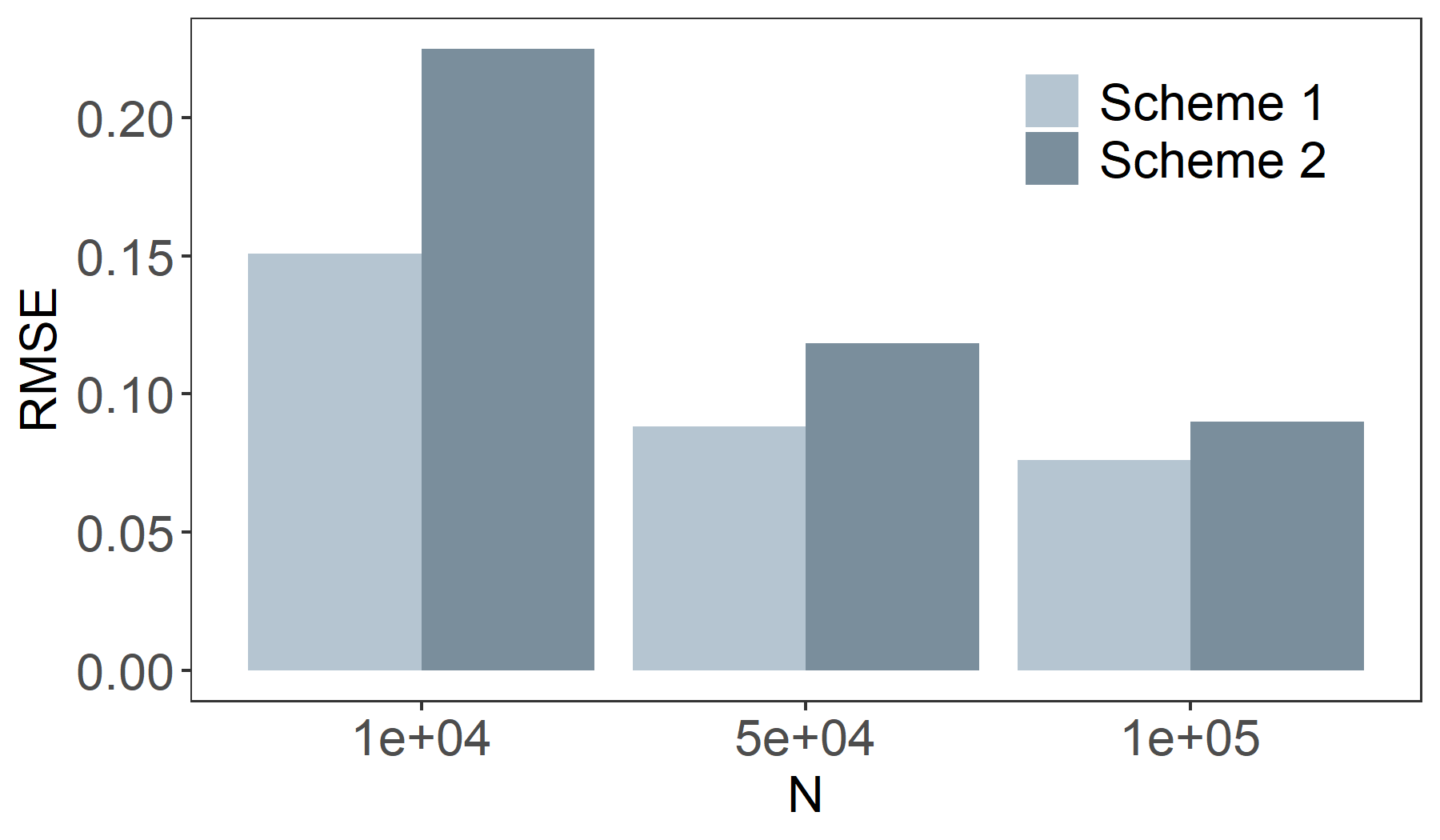}
			\caption{$t(2)$ and $K = \lfloor n^{0.4}\rfloor$.}
		\end{subfigure}
		\begin{subfigure}{0.49\textwidth}
			\includegraphics[width=\textwidth]{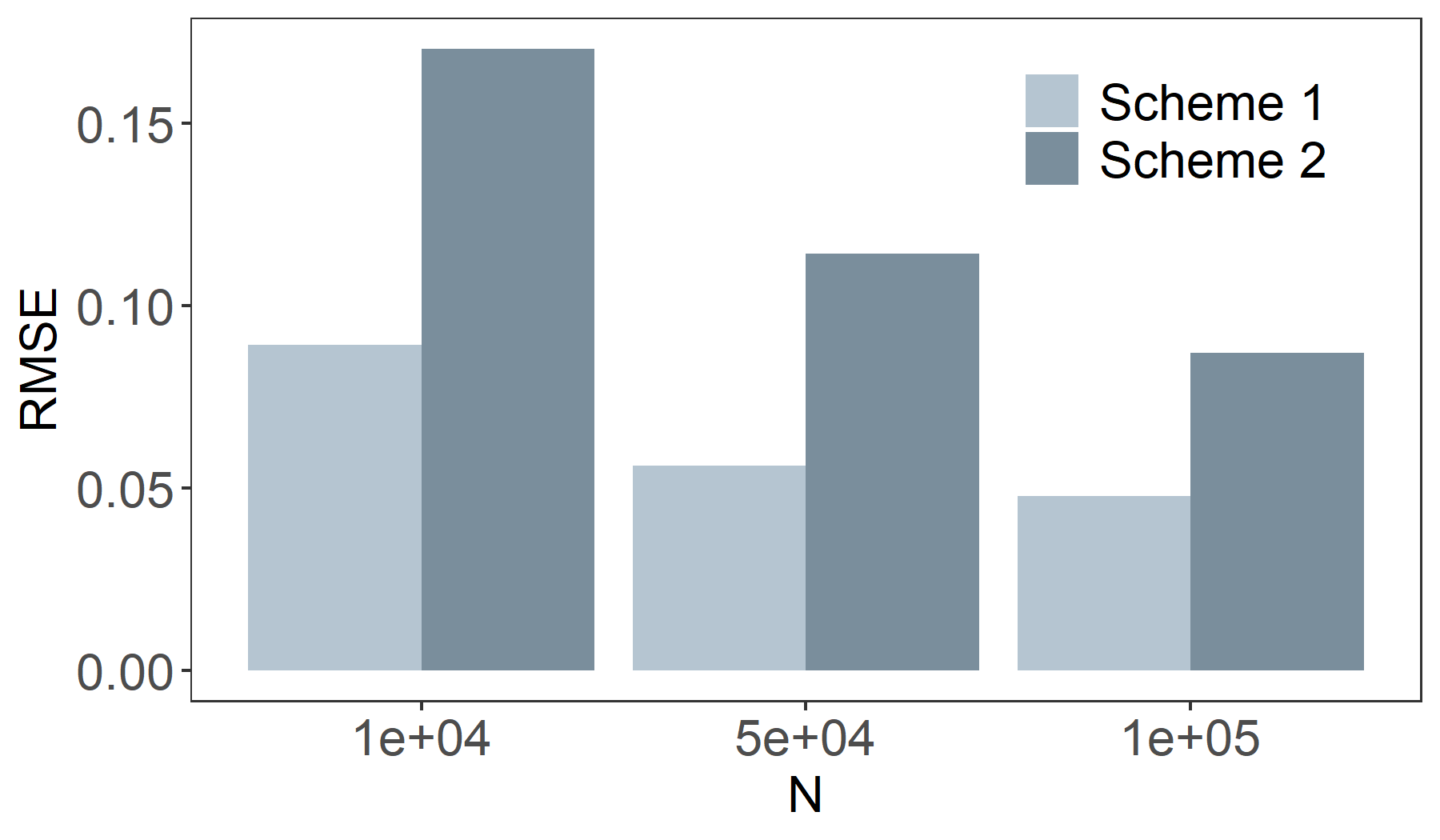}
			\caption{$t(2)$ and $K = \lfloor n^{0.6}\rfloor$.} 
		\end{subfigure} \\[3mm]
		\caption{RMSEs of AML estimators for different threshold selection schemes over 100 replications. We consider different data generating functiions and different number of subsamples $K$ in different plots. The subsample size $n$ is fixed at $n = \lfloor N^{0.5}\rfloor$.}
		\label{fig:threshold}
	\end{figure}

	To evaluate the performance of these two schemes in $u$ selection, we consider to generate simulated dataset from distributions t(1) and  t(2).  The whole sample size $N$ varies from $10^4$ to $ 10^5$. The subsample size $n$ is fixed at $n=\lfloor N^{0.5} \rfloor $. The number of subsamples $K$ is fixed at $K = \lfloor n^{0.4}\rfloor $ and $K = \lfloor n^{0.6}\rfloor$ respectively.  We consider a total of 100 candidate values for $u$, with the corresponding empirical tail probabilities (based on the first subsample) are evenly distributed on the interval [0.005,0.5]. Then we perform the two different schemes to select the optimal value of $u$. These lead to two different values of AML estimator.  Figure \ref{fig:threshold} presents the RMSEs of AML estimators  over 100 replications. As expected,  \textsf{Scheme 1} outperforms  \textsf{Scheme 2} by a noticeable margin. It is better to use \textsf{Scheme 1} for threshold selection. However, since \textsf{Scheme 2} only needs to go through all subsamples once, \textsf{Scheme 2} might be  practically more convenient if the time budget is extremely limited.

	\csection{THE  AIRLINE DATA ANALYSIS}

	To demonstrate the practical usefulness of the proposed method, we present here a real dataset analysis. The dataset is referred to as the airline data. It can be freely downloaded from the website \url{http://stat-computing.org/dataexpo/2009/}. Each record contains detailed information for one particular commercial flight  within the USA,
	from 1987 and 2008. Specifically, a total of 29 variables are included in each record. Among these variables, 13 of them are continuous variables. However, most of these continuous variables suffer from severe missing problems. Therefore, we will only focus on four continuous variables with missing rates less than 10\% for illustration purpose.  They are, respectively, the  \textsf{ActualElapsedTime} (actual elapsed time), \textsf{CRSElapsedTime} (scheduled elapsed time), \textsf{ArrDelay} (arrival delay) and \textsf{DepDelay} (departure delay)  variables.  More details about these variables can be found in the downloading website.

	Our aim in this study is to obtain the 99.99\% quantiles of the underlying distributions that corresponds to these four continuous variables respectively. Note that the original dataset contains about 120 million  (116,525,241) records. It takes up about 12 GB on a hard drive. As a consequence, it is too large to be read into a usual personal computer's memory as a whole. We have to rely on subsampling technique for data analysis.   To achieve the best subsampling efficiency, we need to preprocess the data on the hard drive appropriately. In particular, the physical address of each data line on the hard drive should be handled carefully. Otherwise, the pointer of the data reader cannot be located to the starting position for each data line efficiently.  The preprocessing and subsampling procedures have been developed into an off-the-shelf python module. It is publicly available online at \url{https://github.com/holybadger/clubear}. 
	With the aid of the python module, we first randomly draw a subsample with size $n=10,000$ for a quick review. Based on the subsample, we are able to calculate the kurtosis for each variable.  The detailed results are reported in Table \ref{tab_real} (in Appendix B). From Table \ref{tab_real}, we can see that all the variables have kurtosis larger than 3. While the kurtosis of the \textsf{ArrDelay} and \textsf{DepDelay} variables are larger than 9. This indicates that all the five variables under study have heavier tails than a normal distribution. Moreover, two of them are even more severely heavy-tailed compared with $t(5)$. 
	The outlier detection for these variables becomes imperative and challenging.

\begin{figure}[!ht]
		\centering
		\includegraphics[scale=.48]{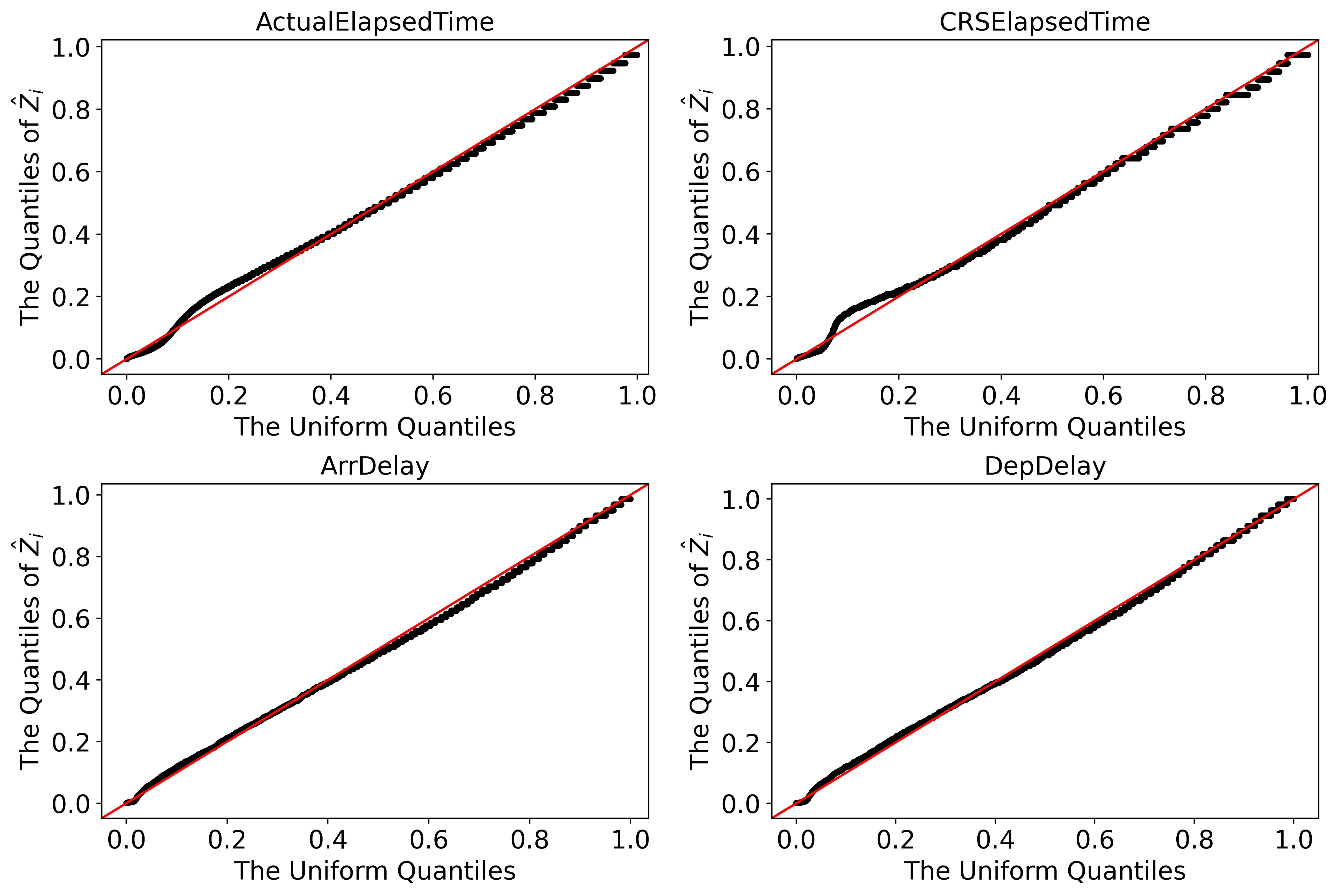}
		\caption{Q-Q plots for four continuous variable we focused. The 45-degree solid line represents the reference line.} 
		\label{qqplot}
	\end{figure}

	To address the issue, we use the proposed subsampling-based method to estimate the  upper and lower bound of  for each variable. {\color{black} The number of subsamples and each subsample size are mainly determined by the memory capacity, expected statistical accuracy and time budgets.}
	In this analysis, we fix $n=10,000$ and $K=100$, which is $O(N^{0.5})$ and $O(n^{0.5})$ respectively. Before the formal data analysis, the original data are appropriately centered so that its tail distribution can be better approximated by the tail probability which specified in Condition (C1). To determine the threshold value $u$, we consider a total of 100 candidate values, with the corresponding  empirical tail probabilities (based on the first subsample) evenly distributed on the interval $[0.005, 0.5]$.  As discussed in Section 3.4, we select the threshold value that minimizes the Cram\'er-von Mises statistic value for each variable.  Specifically, the optimal threshold values for \textsf{ActualElapsedTime}, \textsf{CRSElapsedTime}, \textsf{ArrDelay}  and \textsf{DepDelay} are the 98.5\%, 98.0\%, 99.5\%, 99.5\% empirical quantiles, respectively. To assess the goodness-of-fit, we  employ the quantile-quantile (Q-Q) plot. If the threshold value is selected properly, then the transformed random variables $\hat{Z}_i$s will be expected to be uniformly distributed on $U[0,1]$. Hence, the corresponding Q-Q plot should approximately follow a 45-degree straight line. Figure \ref{qqplot} displays the Q-Q plots corresponding to the selected optimal threshold value for each variable. They all approximately follow a 45-degree straight line. Hence, we conclude that the selected threshold values are satisfactory.

	Given the threshold value, we can calculate $\hat{\gamma}_{k}$ for each subsample and finally obtain the AML estimator  $\hat{\gamma}_{\text{\sc aml}}$. Then we are able to compute the value of the quantile estimator $\hat{q}^{(a)}_{1-\tau}$ according to its formula. The quantile level $\tau$ is fixed to be $10^{-4}$. In other words, for each variable, $99.99\%$ of the normal records are expected to be no more  than $\hat{q}^{(a)}_{1-\tau}$. Hence, the value of $\hat{q}^{(a)}_{1-\tau}$ can be served as an upper bound for each variable. Observations that fall outside the bound will be considered as {\color{black} suspected outliers and be treated separately.} From Table \ref{tab_real}, we observe that, for all variables under study except \textsf{DepDelay}, less than 0.01\% of observations are detected as suspected outliers. However, for \textsf{DepDelay}, the suspected outlier probability becomes 0.011\%,  which is larger than the nominal level 0.01\%. This suggests that for \textsf{DepDelay}, {\color{black} some of those suspected outliers deserve our attention. Their extreme behaviors can hardly be explained by the inherent tail-heaviness of the data distribution.
		Therefore, practically it needs further and more careful analysis to  investigate whether these points are genuine outliers or highly influential observations. 
		
		Lastly, we compare the total time cost between the subsampling based method and  the whole data based method. Note that the airline dataset is too large to be  loaded into memory as a whole. Consequently, if the whole data based estimator is to be computed, it has to be partitioned into many non-overlapping small pieces. Subsequently these small pieces can be processed in a part-by-part manner. 
		For convenience, we refer to each small piece as a subsample and fix the subsample size to be $n=10,000$. This leads to a total of 11,653 non-overlapping subsamples.  Our numerical experiment suggests that it takes about 2.5 seconds on average to finish all the necessary computation jobs  for each subsample. This includes the time costs for data loading, parameter estimation for a total of 100  candidate threshold values, and threshold evaluation according to the discrepancy measure, as developed in Section 3.4.  Consequently, if the whole dataset is to be processed, the total time cost is about $11,653 \times 2.5$ seconds $\approx 8$ hours. In contrast, for our proposed subsampling method, the total time cost  can be significantly reduced to be  $K \times 2.5$ seconds with $K=100$ (less than 5 minutes). A workflow chart in this regard is illustrated in Figure \ref{workflow}. The whole dataset based global estimators (denoted as $\hat{\gamma}_{\text{global}}$) and the subsample based estimators (i.e., $\hat{\gamma}_{\text{AML}}$) for the four interested variables are reported in Table $\ref{tab_real}$.
		We find that the absolute differences between
		$\hat{\gamma}_{\text{global}}$ and $\hat{\gamma}_{\text{AML}}$ are less than $0.12\%$ for all variables. It seems to us that such a difference is tiny and  practically very acceptable for most real applications.

	\csection{CONCLUDING REMARKS} 
	
	In this paper, we develop a subsampling-based method to estimate the extreme value index for heavy-tailed data. The method is developed particularly for massive datasets with memory constraints. By repeatedly subsampling from the whole data, we can estimate the extreme value index consistently. With the help of the  extreme value index estimator, the  corresponding estimations for high-level quantiles and tail probabilities can also be obtained.  The analysis from both simulated and  real datasets  demonstrates the  effectiveness and efficiency of the method.  Moreover, this method does not require  luxury computing devices, such as parallel computing servers. It is more flexible and available for most practitioners with limited computing resources.

	To conclude this paper, we would like to discuss a number of interesting problems for further research. First, 
	it would be worthy to extend our method to multivariate settings. Since the large-scale data in reality are often high-dimensional, the tail dependence  between different variables might be nonnegligible. The way to take this information into consideration  deserves further investigation. Second, recall that in this paper, $X_i$s are assumed to be independent and identically distributed random variables with extreme value index  $\gamma$. Let $\xi$ be an arbitrary but fixed constant. Theoretically speaking, both  variables $X_i$ and $X_i+\xi$ should have the same extreme value index. Accordingly, the interested extreme value index can be estimated by either the original sample $\{X_i\}$ or the shifted one $\{X_i+\xi\}$. Both samples should lead to consistent estimators for $\gamma$. However, our numerical experience suggests that their finite sample performances could be very different. An appropriately shifted sample can yield an estimator substantially more efficient than that of the original one. As a consequence, how to select the optimal shifting parameter should be another interesting topic for future study. Third, apart from uniform subsampling, a series of novel non-uniform subsampling methods have also been developed in recent years \citep{ma2015statistical, wang2018optimal, han2020local, ai2021optimal,wang2021optimal}. For non-uniform sampling, it generally aims to approximate the whole dataset based estimators by selecting informative observations with higher probabilities. Thus, it leads to more efficient estimators with a fixed subsample size.  Developing a similar method for estimating the extreme value index will be a useful and interesting research topic, which is worth further exploration.

	\newpage

	\renewcommand{\theequation}{A.\arabic{equation}}
	\setcounter{equation}{0}
	
	\scsection{APPENDIX A. LEMMAS }
	
	In this section, we present here several useful lemmas for proofs of theorems in Appendix B. To prove the first conclusion in Theorem \ref{th:sub}, the following lemma is needed. Similar lemma was also developed in \cite{wang2009tail}. The difference is that the lemma given in Wang and Tsai (2009) was developed for original data while ours is developed for subsampled data. The details are given below.
	
	\begin{lemma}
		\label{lem1} 
		Under Conditions (C1) and (C2), we have $n_k^u/n=\beta u^{-1/\gamma}\{1+o_p(1)\}$.
		
		\begin{proof}
			The conclusion follows, if we can show that
			$A = \beta^{-1}u^{1/\gamma}n_k^u/n = 1 +o_p(1)$. To this end, it suffices to show that $E(A)=1+o(1)$ and $\text{var}(A)=o(1)$. For the expectation of $A$, let $\alpha_u^*=P(X_{i}>u)$ where $i\in \mathcal{S}_k$ for any $1\leq k \leq K$. Then we have $E(A)=\beta^{-1}u^{1/\gamma}\alpha_u^*$. Note that 
			\begin{align*}
			\alpha_u^*  & =P(X_{i}>u) =E\big\{I(X_{i}>u)\big\}  = E\Big[E\left\{I(X_{i}>u) \big| \mS \right\}\Big] \nonumber \\
			& = E\left\{N^{-1}\sum_{i\in \mS} I(X_i>u)\right\} = P(X>u)=\alpha_u.
			\end{align*}
			Moreover, Condition (C1) implies that $\alpha_u = \beta^{-1}u^{1/\gamma}\{1+o(1)\}$. Therefore, $E(A)=\beta^{-1}u^{1/\gamma}\alpha_u=1+o(1)$. For the variance of $A$, write $U_{i}=I(X_{i}>u)$, then
			$\text{var}(A)=(\beta^{-1}u^{1/\gamma})^2\text{var}(n_k^u/n) = n^{-2}(\beta^{-1}u^{1/\gamma})^2  \text{var}\left( \sum_{i \in \mathcal{S}_k} U_{i} \right)$. Recall that 
			\begin{align*}
			& \quad\ \text{var}\left(\sum_{i \in \mathcal{S}_k} U_{i}\right) =  E\left\{ \text{var}\bigg( \sum_{i \in \mathcal{S}_k} U_{i} \Big|\mS \bigg) \right\} + \text{var} \left\{E\bigg( \sum_{i \in \mathcal{S}_k} U_{i} \Big|\mS \bigg) \right\} \\
			& = E\left\{n N^{-1} \sum_{i \in \mS} \big( U_i - \widebar{U} \big)^2 \right\}
			+ \text{var}\left(nN^{-1}\sum_{i \in \mS} U_i\right) \\ & =n\big(n+N-1\big)N^{-1}\text{var}\big(U\big). 
			\end{align*}
			where $\widebar{U}=N^{-1}\sum_{i \in \mS} U_i$, $U=I(X>u)$.
			Moreover, $\text{var}(U) = P(X>u)\{1-P(X>u)\} \leq P(X>u)$.  Therefore, combining with Condition (C1), we have 
			\begin{align*}
			\text{var}\big(A\big) &  \leq (\beta^{-1}u^{1/\gamma})^2 (n+N-1)(nN)^{-1}P\big(X>u\big)\\ & = \beta^{-1}(un^{-\gamma})^{1/\gamma}(n+N-1)N^{-1}\{1+o(1)\}.
			\end{align*}
			By Condition (C2) and (C3), we know that $n=o(N)$ and $u=o(n^{\gamma})$. Hence, $\text{var}(A)\rightarrow 0$. This completes the proof.
		\end{proof}
	\end{lemma}
	
	To prove the second conclusion in Theorem \ref{th:sub}, we need the following three lemmas. We first develop an Bernstein-type inequality for subsampled data in Lemma \ref{bern}. This lemma can be viewed as the extension of the classical Bernstein inequality \citep{bennett1962probability} but for subsampled data.
	\begin{lemma}
		\label{bern}
		Let $\mathcal{F}_N=\{X_1, \cdots, X_N\}$ be independent and identically distributed bounded random variables such that $EX_i=\mu$ and $|X_i| \leq M$. Subsamples $\{X_1^*, \cdots, X_n^*\}$ are drawn from $\mathcal{F}_N$ independently with replacement. Then for any $t>0$, we have 
		\begin{align*}
		P\bigg(\Big|n^{-1}\sum_{i=1}^n X_{i}^* -\mu \Big|> t \bigg)  \leq  4\exp\left(-\frac{nt^2}{12\phi+4Mt/3}\right)+2\exp\left(-\frac{N\phi^2}{8\psi^2+4M^2\phi/3}\right),
		\end{align*}
		where $\phi=E(X_i^2)$ and $\psi = \text{var}(X_i^2)$.
		\begin{proof}
			Let $\sigma^2=\text{var}(X_i)$. By Bernstein inequality \citep{bennett1962probability}  we know that, for any $t>0$,  $P(\sum_{i=1}^N X_i -N\mu >t)\leq \exp\{-t^2/(2N\sigma^2+2Mt/3)\}$. This is equivalent to
			\begin{align}
			\label{bern1}
			P\Big(\big|\widebar{X} -\mu\big| > t \Big) \leq 2\exp\left(-\frac{Nt^2}{2\sigma^2+2Mt/3}\right).
			\end{align}
			Similarly, note that given full dataset $\mathcal{F}_N$, $X_i^*$s are conditionally independent with $E(X_i^*|\mathcal{F}_N)=N^{-1}\sum_{i=1}^N X_i=\widebar{X}$ and $\text{var}(X^*_i|\mathcal{F}_N)=N^{-1}\sum_{i=1}^N(X_i - \widebar{X})^2 \leq N^{-1}\sum_{i=1}^N X_i^2$, we can also obtain that
			\begin{align}
			\label{berncond}
			P\bigg(\big|n^{-1}\sum_{i=1}^n X_{i}^*- \widebar{X} \big|> t \Big|\mathcal{F}_N \bigg)  \leq  2\exp\bigg(-\frac{nt^2}{2N^{-1}\sum_{i=1}^N X_i^2+2Mt/3}\bigg).
			\end{align}
			Taking the expectation from both sides in  (\ref{berncond}) lead to
			\begin{align}
			\label{berncond2}
			& \quad\ P\Big(\big|n^{-1}\sum_{i=1}^n X_{i}^*- \bar{X} \big| > t \Big) = E\bigg\{P\Big(\big|n^{-1}\sum_{i=1}^n X_{i}^*- \bar{X} \big| > t \Big|\mathcal{F}_N \Big)\bigg\} \nonumber \\
			&  = E\bigg\{P\Big(\big|n^{-1}\sum_{i=1}^n X_{i}^*- \bar{X} \big| > t \Big|\mathcal{F}_N \Big) 
			I\Big(\big|N^{-1}\sum_{i=1}^NX_i^2-\phi\big|<\phi/2\Big) +  \nonumber \\
			& \qquad \qquad   \qquad \qquad  \qquad \quad     P\Big(\big|n^{-1}\sum_{i=1}^n X_{i}^*- \bar{X} \big| > t \Big|\mathcal{F}_N \Big) 
			I\Big(\big|N^{-1}\sum_{i=1}^NX_i^2-\phi\big|>\phi/2\Big) \bigg\}
			\nonumber \\ & \leq  2\exp\left(-\frac{nt^2}{3\phi+2Mt/3}\right) + P\Big(\big|N^{-1}\sum_{i=1}^NX_i^2-\phi\big|>\phi/2\Big). 
			\end{align}
			For the second term on the right side of (\ref{berncond2}), notice that $E(N^{-1}\sum_{i=1}^NX_i^2)=\phi$ and $X_i^2\leq M^2$, applying Bernstein inequality (\ref{bern1}) again and replacing $t$ by $\phi/2$ gives
			\begin{align}
			\label{berncond3}
			P\Big(\Big|N^{-1}\sum_{i=1}^NX_i^2-\phi\Big|>\phi/2\Big) \leq 2\exp\left(-\frac{N\phi^2}{8\psi^2+4M^2\phi/3}\right).
			\end{align}
			Combining (\ref{bern1})-(\ref{berncond3}) and $n\leq N$ and $2\sigma^2<3\phi$ yield that $P(|n^{-1}\sum_{i=1}^n X^*_i -\mu|> 2t) \leq 4 \exp\{-nt^2/(3\phi+2Mt/3)\}+2\exp\{-N\phi^2/(8\psi^2+4M^2\phi/3)\}$. This completes the proof.
		\end{proof}
		
	\end{lemma}
	
	\begin{lemma}
		\label{lem2} 
		For $0< t\leq 1$, we have 
		\begin{align*}
		P\Big(\max_{1\leq k \leq K}\big|\alpha_un/n_k^u -1\big|>t\Big) \leq 8K\exp\Big(-\alpha_unt^2/51\Big)+4K\exp\Big(-3\alpha_uN/28\Big).
		\end{align*}
		\begin{proof} For $0<t\leq1$, we have	
			\begin{align*}
			\MoveEqLeft P\Big(\max_{k}\big|\alpha_un/n_k^u -1\big|>t\Big) \leq KP\Big(\big|\alpha_un/n_k^u -1\big|>t\Big) \\
			& \qquad \leq KP\Big\{\big|n_k^u/(\alpha_un)-1\big|>tn_k^u/(\alpha_un), \big|n_k^u/(\alpha_un)-1\big|\leq t/2\Big\}\\
			& \qquad \qquad \qquad  \qquad \qquad \qquad  \qquad \qquad \qquad\  +  KP\Big\{\big|n_k^u/(\alpha_un)-1\big|>t/2\Big\}\\
			&  \qquad \leq  KP\Big\{\big|n_k^u/(\alpha_un)-1\big|>t(1-t/2)\Big\} +KP\Big\{\big|n_k^u/(\alpha_un)-1\big|>t/2\Big\}\\
			&  \qquad \leq 2KP\Big\{\big|n_k^u/(\alpha_un)-1\big|>t/2\Big\}. 
			\end{align*}
			Notice that $n_k^u/(\alpha_un)= n^{-1}\sum_{i \in \mathcal{S}_k}H_{i}$, where $H_{i}=\alpha_u^{-1}I(X_{i}>u)$ bounded by $M=\alpha_u^{-1}$. According to the proof of Lemma \ref{lem1}, $E(H_{i})=1$ and $\text{var}(H_{i})=\alpha_u^{-1}(1-\alpha_u)\leq \alpha_u^{-1}$. Moreover, $H_{i}^2 = \alpha_u^{-1} H_{i}$ implies that $\phi=E(H_{i}^2)=\alpha_u^{-1}$ and $\psi^2=\text{var}(H_{i}^2)\leq\alpha_u^{-3}$. Combining the results and Lemma \ref{bern} lead to 
			\begin{align*}
			P\Big(\big|n_k^u/(\alpha_un)-1\big|>t/2\Big)  &\leq 4\exp\left(-\frac{nt^2}{48\phi+8Mt/3}\right)+2\exp\left(-\frac{N\phi^2}{8\psi^2+4M^2\phi/3}\right)\\
			&\leq 4\exp\Big(-\alpha_unt^2/51\Big)+2\exp\Big(-3\alpha_uN/28\Big).
			\end{align*}
			Here, the second inequality follows that $48\phi + 8Mt/3 \leq \alpha_u^{-1}(48+8t/3) \leq \alpha_u^{-1}(48+8/3) \leq 51 \alpha_u^{-1}$ and 
			$8\psi^2+4M^2\phi/3 =8\psi^2+4\alpha_u^{-3}/3\leq 8\alpha_u^{-3}+4\alpha_u^{-3}/3=28\alpha_u^{-3}/3$. This completes the proof.				
		\end{proof}	
	\end{lemma}
	
	\begin{lemma}
		\label{lem3} Under Conditions (C1)-(C4), 
		$K^{-1/2} \sum_{k=1}^K  (Q_{1k}^2-1)Q_{2k} \xrightarrow{p} 0$ as $N\rightarrow \infty$.
		\begin{proof}
			This lemma conclusion follows if we can show that, for any $\epsilon >0$, we have	$P\{|K^{-1/2} \sum_{k=1}^K  (Q_{1k}^2-1)Q_{2k}| > \epsilon \}\rightarrow 0$ as $N\rightarrow \infty$. Note that
			\begin{align}
			& \quad\	P\Big\{\big|K^{-1/2} \sum_{k=1}^K  (Q_{1k}^2-1)Q_{2k}\big| > \epsilon   \Big\}  \leq 	P\Big(K^{-1/2} \sum_{k=1}^K  \big|Q_{1k}^2-1\big|\big|Q_{2k}\big| > \epsilon \  \Big) \nonumber \\
			& \leq P\Big( \max_{1\leq k \leq K} \big|Q_{1k}^2-1\big| K^{-1/2} \sum_{k=1}^K \big|Q_{2k}\big| > \epsilon   \Big), 
			\label{lem3:1}
			\end{align}	
			Therefore, the conclusion follows if we are able to show that $ \max_{1\leq k \leq K} |Q_{1k}^2-1|K^{1/2}=o_p(1)$ and  $ K^{-1} \sum_{k=1}^K |Q_{2k}| =O_p(1)$. For the first one, since $Q_{1k}^2$=$\alpha_un/n_k^u$, then by Lemma \ref{lem2} we have for any $\varepsilon>0$ such that $\varepsilon/\sqrt{K}\leq 1$,
			\begin{align*}
			P\Big(\max_{1\leq k \leq K} \big|Q_{1k}^2-1\big|>\varepsilon/\sqrt{K} \Big) \leq 8K\exp\Big(-\alpha_un\varepsilon^2/51K\Big)+4K\exp\Big(-3\alpha_uN/28\Big).
			\end{align*}
			If $\varepsilon/\sqrt{K}>1$, we can still obtain the upper bound by
			\begin{align*}
			\MoveEqLeft  P\Big(\max_{1\leq k \leq K} \big|Q_{1k}^2-1\big|>\varepsilon/\sqrt{K} \Big)  \leq 
			P\Big(\max_{1\leq k \leq K} \big|Q_{1k}^2-1\big|>1\Big)\\
			& \qquad \qquad \qquad \qquad \leq 
			8K\exp\Big(-\alpha_un/51\Big)+4K\exp\Big(-3\alpha_uN/28\Big).
			\end{align*}
			According to Conditions (C1)-(C4), $\alpha_u n$ and $K \rightarrow \infty$ as $N\rightarrow \infty$. 
			Condition (C4) indicates that $ K =o(\alpha_u n)$. Therefore, for any $\varepsilon>0$, $P\{\max_{1\leq k \leq K} |Q_{1k}^2-1|K^{1/2}>\varepsilon\} \rightarrow 0$ as $N\rightarrow \infty$ and thus  $ \max_{1\leq k \leq K} |Q_{1k}^2-1|K^{1/2}=o_p(1)$.
			
			For the second one, note that in Appendix A.1, it has been shown that $E(Q_{2k})=o(1)$ and $\text{var}(Q_{2k})=\gamma^2+o(1)$. Therefore, $E(K^{-1} \sum_{k=1}^K |Q_{2k}|)=E(|Q_{2k}|)\leq E(Q_{2k}^2)^{1/2}=O(1)$, implying that $K^{-1} \sum_{k=1}^K |Q_{2k}|=O_p(1)$. This completes the proof.
		\end{proof}
	\end{lemma}

	\renewcommand{\theequation}{B.\arabic{equation}}
	\setcounter{equation}{0}
	
	\scsection{APPENDIX B. PROOF OF THEOREMS}
	
	In this appendix, we provide the full proof of Theorem \ref{th:sub} and Theorem \ref{th2}.

	\scsubsection{B.1. Proof of Theorem \ref{th:sub}}
	
	\noindent{\it 1. Proof of Conclusion (i).}
	
	Write $\sqrt{n_k^u}(\hat{\gamma}_{k}-\gamma) =(n_k^u)^{-1/2}\sum_{i\in \mathcal{S}_k} \left\{ \log(X_{i}/u) - \gamma \right\}I(X_{i}>u)=Q_{1k}  Q_{2k} $, where $Q_{1k} = (\alpha_u n/n_k^u)^{1/2}$ and $Q_{2k} = (\alpha_u n)^{-1/2}\sum_{i\in \mathcal{S}_k} \left\{ \log(X_{i}/u) - \gamma \right\}I(X_{i}>u)$.
	To prove the result in Theorem \ref{th:sub}, it suffices to show that $Q_{1k} \xrightarrow{p} 1 $ and  $Q_{2k} \xrightarrow{d} N(0, \gamma^2)$. By Lemma \ref{lem1} in Appendix B, we know that $\alpha_u n/n_k^u \xrightarrow{p}1$. This implies that $Q_{1k} \xrightarrow{p}1$. Moreover, $Q_{2k}$ is a normalized sum of independent and identically distributed random variables conditioned on $\mS$. Denote $Y_{i} = \alpha_u^{-1/2} \left\{ \log\big(X_{i}/u\big) - \gamma \right\} I\big(X_{i}>u\big)$. Thus, by the Lindeberg-Feller theorem, the asymptotical normality of $Q_{2k}$ holds if we are able to show that (i)
	$E(Q_{2k}) = o(1)$, (ii) $\text{var}(Q_{2k}) = \gamma^2 +o(1)$ and (iii) $\sum_{i\in \mathcal{S}_k}E(n^{-1}|Y_{i}|^2 1\{n^{-1/2}|Y_{i}|>\epsilon\})=o(1) $ for any $\epsilon>0$. The proof details are given below.

	{\sc Step 1.} We study $E(Q_{2k})$ first. Note that $E(Q_{2k})=E\left\{E(Q_{2k}|\mS)\right\}$. Given full data $\mS$,  the conditional expectation of $Q_{2k}$ equals 
	\begin{align}
	E\big(Q_{2k} \big|\mS\big)& = n^{-1/2} E\left[\sum_{i\in \mathcal{S}_k} \alpha_u^{-1/2}\bigg\{ \log\big(X_{i}/u\big) - \gamma \bigg\} I\big(X_{i}>u\big)\Big|\mS\right] \nonumber \\ 
	& =  n^{1/2}N^{-1} \sum_{i\in \mS} Y_i,  \label{cond_e}
	\end{align}
	Hence,
	$E(Q_{2k})=n^{1/2}E(Y)$ where $Y = \alpha_u^{-1/2} \left\{ \log\big(X/u\big) - \gamma \right\} I\big(X>u\big)$. Next we examine  the expectation of $Y$.
	
	By the definition of $Y$, 
	$ E(Y) =  \alpha_u^{-1/2} \left[E\left\{ \log\big(X/u\big)I\big(X>u\big)\right\} - \gamma P\big(X>u\big)\right]$.
	Note that for the first term, 
	\begin{align}
	E\Big\{ \log\big(X/u\big)I\big(X>u\big)\Big\} 
	& = \int_{0}^{\infty}P\Big\{ \log\big(X/u\big)>t\Big\} dt = \int_{0}^{\infty}P\big(X>ue^t\big) dt \nonumber \\
	& \overset{(a)}{=}\ \int_1^{\infty} P(X>u s) \frac{1}{s} ds 
	=  P(X>u)  \int_1 ^{\infty} \frac{P(X>us)}{P(X>u)}\frac{1}{s}ds  \nonumber \\ 
	&  \overset{(b)}{=}  P(X>u) \int_1 ^{\infty} \frac{L(us)}{L(u)}s^{-1/\gamma - 1}ds,
	\label{th1:pf1}
	\end{align}
	where in the third equality (a), we let $s=e^t$. In the last equality (b), we use the notation $L(u) = u^{1/\gamma} P(X>u)$.
	Then we have $L(us)/L(u) = 1 + C(s^{-\delta}-1)u^{-\delta} + o(u^{-\delta})$.
	Denote $k(s) = C(s^{-\delta}- 1)$ and $\phi(u)=u^{-\delta}$, then
	$L(us)/L(u) = 1 + k(s)\phi(u) + o(\phi(u))$, as $ u \rightarrow \infty$ for each $s>0$.
	We next define $v(s) = s^{-1/\gamma - 1}$. It can be shown that $v(s)$ is integrable. By the Proposition 3.1  in  \cite{smith1987estimating}, we obtain that
	\begin{align}
	\int_1^{\infty}  \frac{L(us)}{L(u)}s^{-1/\gamma - 1}ds  & = \int_1^{\infty} v(s)ds  + \phi(u) \int_1^{\infty} v(s)k(s)ds + o(\phi(u)) \nonumber \\
	& = \gamma + C\left\{(1/\gamma + \delta)^{-1} - \gamma \right\} \phi(u)  + o(\phi(u)). \label{th1:pf2}
	\end{align}
	Plugging the last two equations (\ref{th1:pf1}) and   (\ref{th1:pf2}) into the formula of $E(Y)$ yields that 
	\begin{align*}
	E(Y) 
	& = \alpha_u^{-1/2}\left[P(X>u) \left\{\gamma + C \left(\frac{1}{1/\gamma + \delta}-\gamma \right)\phi(u) + o(\phi(u))\right\} - \gamma P(X>u) \right]\\
	& =  \alpha_u^{-1/2}\beta C\left\{(1/\gamma + \delta)^{-1} - \gamma \right\} u^{-1/\gamma - \delta} + o(1).
	\end{align*}
	Therefore, 
	\begin{align*}
	E(Q_{2k}) = n^{1/2}E(Y) = \beta C \left( \frac{1}{1/\gamma + \delta}- \gamma \right) \Big(nu^{-1/\gamma-2\delta}\Big)^{1/2}\left(\frac{u^{-1/\gamma} }{\alpha_u}\right)^{1/2} +o(1).
	\end{align*}	
	By Condition (C3), we have $nu^{-1/\gamma-2\delta} \rightarrow 0$. And by Condition (C1), we have $\alpha_u^{-1}u^{-1/\gamma}=O(1)$. This implies that $E(Q_{2k})=o(1)$.
	
	{\sc Step 2.} We next study the variance of $Q_{2k}$. 
	Here, $\text{var}(Q_{2k})  =  P_1 +P_2$, where $P_1 = \text{var} \{ E (Q_{2k}|\mS) \}, P_2= E\{ \text{var}(Q_{2k}|\mS)\} $.
	For $P_1$, note that $E\big(Q_{2k} \big|\mS\big)= n^{1/2}N^{-1} \sum_{i\in\mS} Y_i$ by (\ref{cond_e}). Hence, $
	P_1 = nN^{-1}\text{var}\big(Y\big)$.
	For $P_2$, notice that the conditional variance of $Q_{2k}$ equals
	\begin{align*}
	\text{var}\big(Q_{2k} \big|\mS\big)& =n^{-1}\text{var}\left[ \alpha_u^{-1/2}\sum_{i\in\mathcal{S}_k} \left\{ \log\big(X_{i}/u\big) - \gamma \right\} I\big(X_{i}>u\big)\Big|\mS\right] \nonumber \\
	& = N^{-1} \sum_{i\in\mS} \left(Y_i - \widebar{Y}\right)^2.
	\end{align*}
	where $\widebar{Y} = N^{-1}\sum_{i\in\mS} Y_i$. Thus, $P_2 = E \big\{ \text{var}\left(Q_{2k}|\mS\right) \big\}=(N-1)\text{var}\left(Y\right)/N$. Then we have  $\text{var}(Q_{2k}) =P_1+P_2 =  (n+N-1)\text{var}\big(Y\big)/N$. 
	Since by Condition (C2), $(n+N-1)/N \rightarrow 1$ as $n\rightarrow \infty$ and $N \rightarrow \infty$. It suffices to show that $\text{var}(Y) \rightarrow \gamma^2$. Next we exam the variance of $Y$.
	
	According to the result of Step 1, we have $E(Y) =n^{-1/2}E(Q_{2k})= o(1)$. Consequently, we obtain that
	\begin{align}
	\text{var}(Y) & = E\left( Y^2 \right) +o(1)  = \alpha_u^{-1}
	E\Big[\big\{\log(X/u)-\gamma \big\}^2I(X>u)\Big] +o(1) \nonumber \\
	& = \alpha_u^{-1} \bigg\{ \int_0^{\infty}P\left\{ \log^2\left(X/u\right)>t\right\}dt - 2\gamma \int_0^{\infty} P\big\{ \log\big(X/u\big)>t\big\} dt + \nonumber \\
	& \qquad \qquad \qquad \qquad \qquad \qquad \qquad \qquad \qquad \gamma^2 P(X>u)\bigg\}  + o(1). \label{eq:vark}
	\end{align}
	Recall that in Step 1, combining (\ref{th1:pf1}) and (\ref{th1:pf2}) gives that $\int_0^{\infty}P\left\{ \log\left(X/u\right)>t\right\} dt= \gamma \alpha_u+ C\left\{ (1/\gamma + \delta)^{-1}- \gamma \right\} \alpha_u\phi(u) + o(1)$ . 
	Proof in a similar manner, we can also obtain that  
	$\int_0^{\infty}P\left\{ \log^2\left(X/u\right)>t\right\} dt=2\gamma^2 \alpha_u + 2C \{(1/\gamma + \delta)^{-2} - \gamma^2 \}\alpha_u \phi(u) +o(1)$. Noting that $\phi(u) = u^{-\delta}$, 
	substituting the two equations into (\ref{eq:vark}), we  have 
	$\text{var}(Y) = \gamma^2 +o(1)$. 
	Thus, $\text{var}(Y) \rightarrow \gamma^2$ as $u\rightarrow \infty$. 
	
	{\sc Step 3.} 
	We finally check the Lindeberg  condition. According to the proof in step 2, we can show that for any $i\in \mathcal{S}_k$, $E(|Y_{i}|^2)$ converges to $\gamma^2$ .
	Note that for any $\epsilon>0$, $\sum_{i\in \mathcal{S}_k}E(n^{-1}|Y_{i}|^2 1\{n^{-1/2}|Y_{i}|>\epsilon\})=E(|Y_{i}|^2 1\{|Y_{i}|>n^{1/2}\epsilon\})$. Then by the dominated convergence theorem, we have $E(|Y_{i}|^2 1\{|Y_{i}|>n^{1/2}\epsilon\})\rightarrow 0$
	as $n \rightarrow \infty $. This completes the proof.
	
	\noindent{\it 2. Proof of Conclusion (ii).}
	
	Note that  $\sqrt{n_{*}^u}(\hat{\gamma}_{\text{\sc aml}}-\gamma) = ({n_{*}^u}/\alpha_unK)^{1/2} \sqrt{\alpha_u n K}(\hat{\gamma}_{\text{\sc aml}}-\gamma)$. Similar to the poof of Lemma \ref{lem1} in Appendix A, it can be shown that $E({n_{*}^u}/\alpha_unK)=1+o(1)$ and $\text{var}({n_{*}^u}/\alpha_unK)=o(1)$. Therefore,  $({n_{*}^u}/\alpha_unK)^{1/2}\xrightarrow{p} 1$ and it suffices to show that $\sqrt{\alpha_unK}(\hat{\gamma}_{\text{\sc aml}}-\gamma)\xrightarrow{d} N(0,\gamma^2)$.
	
	According to the poof of Conclusion (i), we have 
	$\sqrt{\alpha_u n}(\hat{\gamma}_{k}-\gamma) =(\alpha_u n/n_k^u)(\alpha_u n)^{-1/2}\\\sum_{i\in \mathcal{S}_k} \left\{ \log(X_{i}/u) - \gamma \right\}I(X_{i}>u)= Q_{1k}^2Q_{2k}$, where $Q_{1k} = (\alpha_u n/n_k^u)^{1/2}$ and $Q_{2k}= (\alpha_u n)^{-1/2}\sum_{i\in \mathcal{S}_k} \left\{ \log(X_{i}/u) - \gamma \right\}I(X_{i}>u)$, following the same definition in the proof of Theorem 1. Hence, $\sqrt{\alpha_u n}(\hat{\gamma}_{k}-\gamma) = Q_{1k}^2Q_{2k}=Q_{2k}+(Q_{1k}^2-1)Q_{2k}$. Therefore, for the averaged estimator $\hat{\gamma}_{\text{\sc aml}}$, we have 
	\begin{align}
	\label{pf2_1}
	\sqrt{\alpha_u n K}(\hat{\gamma}_{\text{\sc aml}}-\gamma)& = K^{-1/2} \sum_{k=1}^K \sqrt{\alpha_u n}(\hat{\gamma}_{k}-\gamma) = K^{-1/2} \sum_{k=1}^K \Big\{ Q_{2k}+\big(Q_{1k}^2-1\big)Q_{2k} \Big\} \nonumber \\
	& = K^{-1/2}\sum_{k=1}^K  Q_{2k} +  K^{-1/2} \sum_{k=1}^K  \big(Q_{1k}^2-1\big)Q_{2k}.
	\end{align}
	For the first term on the right hand side of (\ref{pf2_1}), it can be proven that under Conditions (C1)-(C4), $E(K^{-1/2}\sum_{k=1}^K  Q_{2k})=(nK)^{1/2}E(Y)=o(1)$ 
	and $\text{var}(K^{-1/2}\sum_{k=1}^K  Q_{2k})=nN^{-1}\text{var}(Y)+(NK-1)\text{var}(Y)/NK=\gamma^2+o(1)$, where $Y$ is defined in Appendix A.1 The rest of proof is simialr to that of $Q_{2k} \xrightarrow{d} N(0,\gamma^2)$ in Appendix A.1 and finally we can conclude that $ K^{-1/2}\sum_{k=1}^K  Q_{2k}  \xrightarrow{d} N(0,\gamma^2)$.    For the second term, Lemma \ref{lem3} in Appendix B indicates that 
	$K^{-1/2}\sum_{k=1}^K  (Q_{1k}^2-1)Q_{2k} \xrightarrow{p} 0$ as $N\rightarrow \infty$. Combining the results together yields that $\sqrt{\alpha_u n K}(\hat{\gamma}_{\text{\sc aml}}-\gamma) \xrightarrow{d} N(0,\gamma^2)$, which completes the proof.

	\scsubsection{B.2. Proof of Theorem \ref{th2}}
	
	\noindent{\it 1. Proof of Conclusion (i).}
	
	The  conclusion (i) follows if we can prove (a) {\color{black}${q}_{1-\tau}^{(a)}/{q_{1-\tau}} \rightarrow 1 $} and (b) $\hat{q}_{1-\tau}^{(a)}/q_{1-\tau}^{(a)} \xrightarrow{p} 1$. For the first part, according to the definition of $q_{1-\tau}$, we have $\tau = \beta q_{1-\tau}^{-1/\gamma}\{1+C q_{1-\tau}^{-\delta}+o(q_{1-\tau}^{-\delta})\}$. Moreover, since $q_{1-\tau}^{(a)} = u\left(\alpha_u/\tau\right)^{\gamma}$, we can also obtain that $\tau=(q_{1-\tau}^{(a)})^{-1/\gamma}u^{1/\gamma}\alpha_u$.
	Combining the two equations together gives 
	$
	\beta q_{1-\tau}^{-1/\gamma}\{1+C q_{1-\tau}^{-\delta}+o(q_{1-\tau}^{-\delta})\} = (q_{1-\tau}^{(a)})^{-1/\gamma}u^{1/\gamma} \beta u^{-1/\gamma}\{1+Cu^{-\delta}+o(u^{-\delta})\}
	$. Rearrange it yields that
	\begin{align}
	\label{pf2}
	(q^{(a)}_{1-\tau}/q_{1-\tau})^{-1/\gamma} = \frac{1+Cq_{1-\tau}^{-\delta}+o(q_{1-\tau}^{-1/\delta})}{1+Cu^{-\delta}+o(u^{-\delta})} .
	\end{align}
	Condition (C3) assumes that $u\rightarrow \infty$ and thus $\alpha_u \rightarrow 0$. Hence, $\tau \rightarrow 0$ and then $q_{1-\tau} \rightarrow \infty$. As a consequence, the right-hand side of (\ref{pf2}) converges to 1, implying that {\color{black}${q}_{1-\tau}^{(a)}/{q_{1-\tau}} \rightarrow 1 $}.
	
	As for the second part, note that  $\hat{q}_{1-\tau}^{(a)} = u\left(\hat{\alpha}_u/\tau\right)^{\hat{\gamma}_{\text{\sc aml}}}$ and ${q}_{1-\tau}^{(a)} = u\left({\alpha}_u/\tau\right)^{\gamma}$, we have 	
	\begin{align}
	\label{pf2:2}
	\log(\hat{q}_{1-\tau}^{(a)}/q_{1-\tau}^{(a)})
	& ={\hat{\gamma}_{\text{\sc aml}}}  \log(\hat{\alpha}_u/\tau) -  \gamma \log(\alpha_u/\tau) \nonumber \\
	& =  ({\hat{\gamma}_{\text{\sc aml}}}-\gamma)( \log \alpha_u - \log \tau) + {\hat{\gamma}_{\text{\sc aml}}}(\log \hat{\alpha}_u- \log \alpha_u).
	\end{align}
	The first term in (\ref{pf2:2}) can be rewritten as  $ ({\hat{\gamma}_{\text{\sc aml}}}-\gamma)( \log \alpha_u - \log \tau)=\sqrt{n_{*}^u}({\hat{\gamma}_{\text{\sc aml}}}-\gamma) (n_{*}^u)^{-1/2}\log (\alpha_u/\tau) $. Here, $\sqrt{n_{*}^u}({\hat{\gamma}_{\text{\sc aml}}}-\gamma)=O_p(1)$
	by Theorem \ref{th:sub}. According to the proof of Lemma 1, we are also able to show that $n_{*}^u=nK\alpha_u\{1+o_p(1)\}$.
	Since we assume that  $\log(\alpha_u/\tau)=o(\sqrt{nK\alpha_u})$, we have  $(n_{*}^u)^{-1/2}\log (\alpha_u/\tau) =o_p(1)$. Therefore, the first term equals to $o_p(1)$.
	
	While for the second term, following an approach similar to that in the proof of Lemma \ref{lem1}, it can be verified that $\hat{\alpha}_u/\alpha_u \xrightarrow{p} 1$ and thus $\log\hat{\alpha}_u - \log \alpha_u \xrightarrow{p} 0$. Since Theorem \ref{th:sub} implies that $\hat{\gamma}_{\text{\sc aml}}$ is a consistent estimator of $\gamma$, we can conclude that 
	${\hat{\gamma}_{\text{\sc aml}}}(\log \hat{\alpha}_u- \log \alpha_u) = o_p(1)$. 
	Combining these two terms together, we have
	$\log(\hat{q}_{1-\tau}^{(a)}/q_{1-\tau}^{(a)}) \xrightarrow{p} 0$. This means that $\hat{q}_{1-\tau}^{(a)}/q_{1-\tau}^{(a)} \xrightarrow{p} 1$.
	The proof is completed.
	
	\noindent{\it 2. Proof of Conclusion (ii).}
	
	Note that $\hat{\tau}=P(X_i > \hat{q}^{(a)}_{1-\tau})=\beta (\hat{q}^{(a)}_{1-\tau})^{-1/\gamma}[1+C(\hat{q}^{(a)}_{1-\tau})^{-\delta}+o\{(\hat{q}^{(a)}_{1-\tau})^{-\delta}\}]$ and $\tau = \beta q_{1-\tau}^{-1/\gamma}\{1+C q_{1-\tau}^{-\delta}+o(q_{1-\tau}^{-\delta})\}$. We have
	\begin{align*}
	\hat{\tau}/\tau = (\hat{q}^{(a)}_{1-\tau}/q_{1-\tau})^{-1/\gamma}  \frac{1+C(\hat{q}^{(a)}_{1-\tau})^{-\delta}+o\{(\hat{q}^{(a)}_{1-\tau})^{-\delta}\}}{1+C q_{1-\tau}^{-\delta}+o(q_{1-\tau}^{-\delta})}.
	\end{align*}
	Since we have shown that $\hat{q}^{(a)}_{1-\tau}/{q}_{1-\tau} \xrightarrow{p} 1$ in the first conclusion, $\hat{\tau}/\tau \xrightarrow{p}1$ can be proven.

	\newpage
	\normalem
	\bibliographystyle{apalike}
	\renewcommand{\section}[2]{}%

	\scsection{REFERENCES}
	\bibliography{Reference}

	\begin{table}[!htp]
		\label{tab1}
		\caption{Simulation results for Example 1 with 1,000 replications. The numerical performance are evaluated for different coefficients $C_K$ (i.e., different numbers of subsamples $K$) and the whole sample sizes $N$. The corresponding subsample sizes $n$ and quantile levels $1-\tau$ (induce the threshold value) are shown in the table. Apart from three measurements (RMSE, ECP and RA) defined in Section 3.2, the average of the total exceedance size ($n_{*}^u$),  bias (Bias) and standard deviation (SD) of $\hat{\gamma}_{\text{\sc aml}}$ are also reported. The target ECP is 95\%.} 
		\centering 
		\vspace{0.25 cm}
		\begin{tabular}{crrrc|rrrrcr}
			\hline
			\hline
			\specialrule{0em}{2pt}{2pt}
			\multirow{2}{*}{$C_K$}&\multicolumn{1}{c}{$N$}&\multicolumn{1}{c}{\multirow{2}{*}{$n$}} &\multicolumn{1}{c}{{$1-\tau$}} &\multirow{2}{*}{$K$}  & \multicolumn{1}{c}{\multirow{2}{*}{$n_{*}^u$}} & Bias & \multicolumn{1}{c}{SD} & \multicolumn{1}{c}{RMSE}&ECP & \multicolumn{1}{c}{RA}\\
			\cline{7-9}	
			\\[-2.2em] &&&&&&&&&& \\
			&\multicolumn{1}{c}{($\times 10^5$)}& & \multicolumn{1}{c}{($\%$)} & &  & \multicolumn{3}{c}{($\times 10^{-2}$)}&(\%) & (\%)
			\\   \specialrule{0em}{1pt}{1pt}
			\hline \specialrule{0em}{3pt}{3pt}
			\multicolumn{11}{c}{{\sc Case 1: }{\it t}{\sc(1) distribution}}	\\ \specialrule{0em}{3pt}{3pt}
			0.3 & $1.0$ & 316 &89.1 &3 & 103.6 &2.76&9.57 &9.96 &96.6 &47.1\\
			& $5.0 $ & 707& 92.0 & 3&169.3&1.34&
			7.86&7.97&95.6&36.2\\
			&   $10.0 $& 1,000 &93.0&3&209.6 &1.03&
			6.87&6.95&95.7&30.6\\
			&   $50.0$& 2,236& 94.9 &4 & 456.9&1.19&
			4.52&4.60&96.5&18.1\\
			&&&&&&&&&& \\
			0.5& $1.0$ & 316 &89.1 &6 & 206.4 &2.47 &7.28 &7.69 &94.9 &36.4\\
			& $5.0 $ & 707& 92.0 & 8&452.3&1.21& 4.62&4.77&96.1&21.0\\
			&   $10.0 $& 1,000 &93.0&9&630.4 &1.11&
			4.01&4.16&94.5&17.6\\
			&   $50.0$& 2,236& 94.9 &13 & 1,482.7&0.43&
			2.57&2.50&95.0&10.4\\
			&&&&&&&&&& \\
			0.7& $1.0$ & 316 &89.1 &14& 482.2 &2.78 &4.61 &5.38 &93.1 &23.5\\
			& $5.0 $ & 707& 92.0 & 21& 1,188.3 &1.53& 3.00&3.36&92.0&13.6\\
			&   $10.0 $& 1,000 &93.0&25&1,751.9 &1.10&
			2.34&2.58&94.0&10.5\\
			&   $50.0$& 2,236& 94.9 &36 &4,105.0&0.54&
			1.56&1.65&94.6&6.4\\
			
			\specialrule{0em}{3pt}{3pt}
			\multicolumn{11}{c}{{\sc Case 2: }{\it t}{\sc(2) distribution}}	\\ \specialrule{0em}{3pt}{3pt}
			0.3& $1.0$ & 316 & 95.9 &2 & 25.9  &3.02 &
			10.86 &11.27 &94.5&104.0\\
			& $5.0$ & 707& 97.4& 2& 36.8  &1.74&
			8.50&8.68&95.5&71.3\\
			&   $10.0$& 1,000 &97.8 &2& 44.4  &1.41&
			8.04&8.16&94.7&59.3\\
			&   $50.0$& 2,236& 98.6 &3 &93.8&1.14&
			5.22&5.35&95.4&30.0 \\
			&&&&&&&&&& \\
			0.5& $1.0$ & 316 & 95.9 &4 & 52.2 &3.34 &
			7.52&8.23 &95.6 &61.1\\
			& $5.0$ & 707& 97.4& 5&91.8  &2.25&
			5.58&6.02&95.0&36.7\\
			&   $10.0$& 1,000 &97.8 &5& 110.2  &1.66&
			4.89&5.16&95.4&32.9\\
			&   $50.0$& 2,236& 98.6 &6 &187.8&1.21&
			3.85&4.03&94.2&21.4 \\
			&&&&&&&&&& \\
			0.7& $1.0$ & 316 & 95.9 &7& 89.9  &2.98 &
			5.62 &6.37&94.8 &45.2\\
			& $5.0$ & 707& 97.4& 9& 165.7 &1.98&
			4.01&4.47&95.4&27.0\\
			&   $10.0$& 1,000 &97.8 &11& 241.6 &1.77&
			3.31&3.75&94.0&20.3\\
			&   $50.0$& 2,236& 98.6 &14 & 439.0&1.15&
			2.45&2.71&93.3&13.8\\
			\specialrule{0em}{1pt}{1pt}
			\hline
		\end{tabular}
	\end{table}

	\begin{table}[!htp]
		\label{tab2}
		\caption{ Simulation results for Example 2 with 1,000 replications. In this simulation, we set $\delta=5$. The numerical performance are evaluated for different coefficients $C_K$ (i.e., different numbers of subsamples $K$) and the whole sample sizes $N$. The corresponding subsample sizes $n$ and quantile levels $1-\tau$ (induce the threshold value) are shown in the table. Apart from three measurements (RMSE, ECP and RA) defined in Section 3.2, the average of the total exceedance size ($n_{*}^u$),  bias (Bias) and standard deviation (SD) of $\hat{\gamma}_{\text{\sc aml}}$ are also reported. The target ECP is 95\%. } 
		\centering 
		\vspace{0.25 cm}
		\begin{tabular}{crrrc|rrrccr}
			\hline
			\hline
			\specialrule{0em}{2pt}{2pt}
			\multirow{2}{*}{$C_K$}&\multicolumn{1}{c}{$N$}&\multicolumn{1}{c}{\multirow{2}{*}{$n$}} &\multicolumn{1}{c}{{$1-\tau$}} &\multirow{2}{*}{$K$}  & \multicolumn{1}{c}{\multirow{2}{*}{$n_{*}^u$}} & Bias & \multicolumn{1}{c}{SD} & \multicolumn{1}{c}{RMSE}&ECP & \multicolumn{1}{c}{RA}\\ \cline{7-9}	
			\\[-2.2em] &&&&&&&&&& \\
			&\multicolumn{1}{c}{($\times 10^5$)}& & \multicolumn{1}{c}{($\%$)} & &  & \multicolumn{3}{c}{($\times 10^{-2}$)}&(\%) & (\%)
			\\   \specialrule{0em}{1pt}{1pt}
			\hline \specialrule{0em}{3pt}{3pt}
			\multicolumn{11}{c}{{\sc Case 1: Pareto(2,1) distribution}}	\\ \specialrule{0em}{3pt}{3pt}
			0.3& $1.0$ & 316 &68.4 &4 & 399.3 &0.01 &5.02 &5.02 &94.6 &32.9\\
			& $5.0 $ & 707& 73.1& 5& 950.8&0.08& 3.33&3.34&94.4&19.3\\
			&   $10.0 $& 1,000 &74.9&5&1,255.1 &0.01&
			2.76&2.76&95.7&15.7\\
			&   $50.0$& 2,236&78.6 &6 &2,873.0&0.01&
			1.82&1.82&95.9&10.1\\
			&&&&&&&&&& \\
			0.5& $1.0$ & 316 &68.4 &11 & 1,098.0 &0.13 &3.04 &3.04 &95.5 &20.1\\
			& $5.0 $ & 707& 73.1& 15& 2,853.4&0.07& 1.88&1.88&94.7&11.3\\
			&   $10.0 $& 1,000 &74.9&17&4,266.7 &-0.06&
			1.56&1.56&94.7&9.1\\
			&   $50.0$& 2,236&78.6 &24 &11,484.4&-0.00&
			0.94&0.94&94.6&5.2\\
			&&&&&&&&&& \\
			0.7& $1.0$ & 316 &68.4 &28 &2,796.5 &-0.02 &2.04 &2.04 &92.2 &15.1\\
			& $5.0 $ & 707& 73.1& 45&8,558.0&0.05&1.14&1.14&93.7&7.7\\
			&   $10.0 $& 1,000 &74.9&56&14,053.9 &0.04&
			0.90&0.90&92.9&5.8\\
			&   $50.0$& 2,236&78.6 &89 &42,581.1&-0.03&
			0.52&0.52&92.2&3.1\\
			\specialrule{0em}{3pt}{3pt}
			\multicolumn{11}{c}{{\sc Case 2: Pareto(2,2) distribution}}	\\ \specialrule{0em}{3pt}{3pt}
			0.3& $1.0$ & 316 & 85.3 &3 & 138.9  &-0.12 &
			4.30 &4.30 &93.2 &52.5\\
			& $5.0$ & 707& 88.8& 4& 316.3  &-0.04&
			2.89&2.89&94.2&30.4\\
			&   $10.0$& 1,000 &90.0 &4&399.7  &0.23&
			2.58&2.59&94.7&24.7\\
			&   $50.0$& 2,236& 92.4 &5 &849.3&-0.02&
			1.69&1.69&95.5&15.7 \\
			&&&&&&&&&& \\
			0.5& $1.0$ & 316 & 85.3 &7 &324.8  &-0.11 &
			2.83 &2.83&94.1 &30.9\\
			& $5.0$ & 707& 88.8& 10&790.0 &-0.04&
			1.69&1.69&95.6&17.4\\
			&   $10.0$& 1,000 &90.0 &11& 1,099.9 &-0.02&
			1.54&1.54&95.0&14.9\\
			&   $50.0$& 2,236& 92.4 &15 &2,550.0&-0.00&
			1.01&1.01&94.9&8.9 \\
			&&&&&&&&&& \\
			0.7& $1.0$ & 316 & 85.3&17 &786.7  &-0.06 &
			1.88 &1.88 &93.9 &20.6\\
			& $5.0$ & 707& 88.8& 26& 2,056.1&-0.02&
			1.13&1.13&94.7&11.3\\
			&   $10.0$& 1,000 &90.0 &31& 3,098.7 &-0.02&
			0.92&0.92&94.2&9.0\\
			&   $50.0$& 2,236& 92.4 &47 &7,986.2&0.00&
			0.56&0.56&95.5&5.0 \\
			\specialrule{0em}{1pt}{1pt}
			\hline
		\end{tabular}
	\end{table}

	\begin{table}[!htp]
		\label{tab4}
		\caption{ Simulation results for Example 3 with 1,000 replications. The numerical performance are evaluated for different coefficients $C_K$ (i.e., different numbers of subsamples $K$) and the whole sample sizes $N$. The corresponding subsample sizes $n$ and quantile levels $1-\tau$ (induce the threshold value) are shown in the table. Apart from three measurements (RMSE, ECP and RA) defined in Section 3.2, the average of the total exceedance size ($n_{*}^u$),  bias (Bias) and standard deviation (SD) of $\hat{\gamma}_{\text{\sc aml}}$ are also reported. The target ECP is 95\%.} 
		\centering 
		\vspace{0.25 cm}
		\begin{tabular}{crrrc|rrrrcr}
			\hline
			\hline
			\specialrule{0em}{2pt}{2pt}
			\multirow{2}{*}{$C_K$}&\multicolumn{1}{c}{$N$}&\multicolumn{1}{c}{\multirow{2}{*}{$n$}} &\multicolumn{1}{c}{{$1-\tau$}} &\multirow{2}{*}{$K$}  & \multicolumn{1}{c}{\multirow{2}{*}{$n_{*}^u$}} & Bias & \multicolumn{1}{c}{SD} & \multicolumn{1}{c}{RMSE}&ECP & \multicolumn{1}{c}{RA}\\  \cline{7-9}	
			\\[-2.2em] &&&&&&&&&& \\
			&\multicolumn{1}{c}{($\times 10^5$)}& & \multicolumn{1}{c}{($\%$)} & &  & \multicolumn{3}{c}{($\times 10^{-2}$)}&(\%) & (\%)
			\\  [1pt] \specialrule{0em}{1pt}{1pt}
			\hline \specialrule{0em}{3pt}{3pt}
			\multicolumn{11}{c}{{\sc Case 1: Fr\'echet(1) distribution}}	\\ \specialrule{0em}{3pt}{3pt}
			0.3& $1.0$ & 316 &95.9 &2 & 26.0 & 0.25& 21.20& 21.20&91.4 &124.0\\
			& $5.0 $ & 707& 97.4 & 2& 36.8 &0.24& 17.03&17.03&93.4&75.4\\
			&   $10.0 $& 1,000 &97.8&2&44.1 &0.88&
			15.19&15.21&93.9&56.2\\
			&   $50.0$& 2,236& 98.6 &3 &93.8&0.09&
			10.84&10.84&93.1&32.2\\
			&&&&&&&&&& \\
			0.5& $1.0$ & 316 &95.9 &4 & 51.5 &1.76&14.79 &14.90 &93.9 &67.3\\
			& $5.0 $ & 707& 97.4 & 5& 92.4 &1.17& 10.71&10.78&95.1&37.0\\
			&   $10.0 $& 1,000 &97.8&5&109.8 &0.45&
			9.95&9.96&94.1&34.5\\
			&   $50.0$& 2,236& 98.6 &6 & 188.3&0.33&
			7.38&7.39&95.1&21.1\\
			&&&&&&&&&& \\
			0.7& $1.0$ & 316 &95.9 &7 &90.5 &1.39&11.18 &11.27 &94.1 &47.6\\
			& $5.0 $ & 707& 97.4 & 9&165.2 &0.86& 8.16&8.20&95.2&28.7\\
			&   $10.0 $& 1,000 &97.8&11&242.3&0.60&
			6.75&6.77&94.2&22.2\\
			&   $50.0$& 2,236& 98.6 &14 &438.3&0.33&
			4.74&4.75&95.0&13.7\\
			
			\specialrule{0em}{3pt}{3pt}
			\multicolumn{11}{c}{{\sc Case 2: Fr\'echet(2) distribution}}	\\ \specialrule{0em}{3pt}{3pt}
			0.3& $1.0$ & 316 &95.9 &2 & 26.0 & 0.13& 10.60& 10.60&91.4 &124.0\\
			& $5.0 $ & 707& 97.4 & 2& 36.8&0.12& 8.51&8.52&93.4&75.4\\
			&   $10.0 $& 1,000 &97.8&2&44.1 &0.44&
			7.59&7.61&93.9&56.2\\
			&   $50.0$& 2,236& 98.6 &3 &93.8&0.04&
			5.42&5.42&93.1&32.2\\
			&&&&&&&&&& \\
			0.5& $1.0$ & 316 &95.9 &4 & 51.5 &0.88&7.40 &7.45&93.9 &67.3\\
			& $5.0 $ & 707& 97.4 & 5& 92.4 &0.59& 5.36&5.39&95.1&37.0\\
			&   $10.0 $& 1,000 &97.8&5&109.8 &0.23&
			4.98&4.98&94.1&34.5\\
			&   $50.0$& 2,236& 98.6 &6 & 188.3&0.16&
			3.69&3.70&95.1&21.1\\
			&&&&&&&&&& \\
			0.7& $1.0$ & 316 &95.9 &7 &90.5 &0.70 &5.59 &5.64 &94.1 &47.6\\
			& $5.0 $ & 707& 97.4 & 9&165.2 &0.43& 4.08&4.10&95.2&28.7\\
			&   $10.0 $& 1,000 &97.8&11&242.3&0.30&
			3.37&3.39&94.2&22.2\\
			&   $50.0$& 2,236& 98.6 &14 & 438.3&0.17&
			2.37&2.37&95.0&13.7\\
			\specialrule{0em}{1pt}{1pt}
			\hline
		\end{tabular}
	\end{table}

	\begin{table}[!htp]
		\label{tab5}
		\caption{ Simulation results for Examples 4 and 5 with 1,000 replications. The numerical performance are evaluated for different coefficients $C_K$ (i.e., different numbers of subsamples $K$) and the whole sample sizes $N$. The corresponding subsample sizes $n$ and quantile levels $1-\tau$ (induce the threshold value) are shown in the table. 
			Apart from three measurements (RMSE, ECP and RA) defined in Section 3.2, the average of the total exceedance size ($n_{*}^u$),  bias (Bias) and standard deviation (SD) of $\hat{\gamma}_{\text{\sc aml}}$ are also reported. The target ECP is 95\%. } 
		\centering 
		\vspace{0.25 cm}
		\begin{tabular}{crrrc|rrrrcr}
			\hline
			\hline
			\specialrule{0em}{2pt}{2pt}
			\multirow{2}{*}{$C_K$}&\multicolumn{1}{c}{$N$}&\multicolumn{1}{c}{\multirow{2}{*}{$n$}} &\multicolumn{1}{c}{{$1-\tau$}} &\multirow{2}{*}{$K$}  & \multicolumn{1}{c}{\multirow{2}{*}{$n_{*}^u$}} & Bias & \multicolumn{1}{c}{SD} & \multicolumn{1}{c}{RMSE}&ECP & \multicolumn{1}{c}{RA}\\
			\cline{7-9}	
			\\[-2.2em] &&&&&&&&&& \\
			&\multicolumn{1}{c}{($\times 10^7$)}& & \multicolumn{1}{c}{($\%$)} & &  & \multicolumn{3}{c}{($\times 10^{-2}$)}&(\%) & (\%)
			\\   \specialrule{0em}{1pt}{1pt}
			\hline \specialrule{0em}{3pt}{3pt}
			\multicolumn{11}{c}{{\sc Example 4  }}	\\
			\specialrule{0em}{3pt}{3pt}
			0.3 & 1.0 & 3,162 & 99.4 & 3 & 61.8 & -0.35 & 12.39 & 12.39 & 94.6 & 28.3 \\
			& 4.0 & 6,324 & 99.6 & 3 & 79.7& 0.70 & 11.81 & 11.83 & 93.9 & 20.6 \\
			& 7.0 & 8,366 & 99.6 & 3 & 88.8 &  -0.18 & 10.66 & 10.66 & 94.2 & 17.2 \\
			& 10.0 & 10,000 & 99.7 & 3 & 95.0 & -0.57 & 9.99 & 10.01 & 94.3 & 15.8 \\
			\specialrule{0em}{3pt}{3pt}
			0.4 & 1.0 & 3,162 & 99.4 & 5 & 102.5 & 0.15 &10.41 & 10.41 & 93.8 & 22.4 \\
			& 4.0 & 6,324 & 99.6 & 5 & 132.7& 0.55 & 8.91 & 8.93 & 93.8 & 15.6 \\
			& 7.0 & 8,366 & 99.6 & 6 & 178.9 & 0.15 & 7.70 & 7.70 & 93.7 & 12.3 \\
			& 10.0 & 10,000 & 99.7 & 6 & 188.9 & -0.10 & 7.34 & 7.34 & 94.4 & 11.3 \\
			\specialrule{0em}{3pt}{3pt}
			0.5 & 1.0 & 3,162 & 99.4 & 7 & 144.1 & 0.36 & 8.53 & 8.54 & 94.3 & 17.7 \\
			& 4.0 & 6,324 & 99.6 & 8 & 212.5 & -0.14 & 6.91 & 6.91 & 94.7 & 12.1 \\
			& 7.0 & 8,366 & 99.6 & 9 & 266.2 &  0.29 & 6.28 & 6.29 & 94.2 & 10.2\\
			& 10.0 & 10,000 & 99.7 & 10 & 315.7 & 0.23 & 5.49 & 5.50& 95.7 & 8.5 \\
			\specialrule{0em}{3pt}{3pt}
			
			\multicolumn{11}{c}{{\sc Example 5}}\\
			\specialrule{0em}{3pt}{3pt}
			0.3 & 1.0 & 3,162 & 99.4 & 3 & 61.8 & 0.54 & 13.64 & 13.65 & 93.5 & 28.2 \\
			& 4.0 & 6,324 & 99.6 & 3 & 79.8& 0.75 & 11.72 & 11.74 & 94.5 & 20.3 \\
			& 7.0 & 8,366 & 99.6 & 3 & 88.5 &  0.07 & 10.73 & 10.73 & 94.9& 17.9 \\
			& 10.0 & 10,000 & 99.7 & 3 & 95.5 & -0.03 & 11.05 & 11.05 & 92.3 & 17.0 \\
			\specialrule{0em}{3pt}{3pt}
			0.4 & 1.0 & 3,162 & 99.4 & 5 & 102.6 & -0.24 &10.29 & 10.29 & 93.6 & 21.9 \\
			& 4.0 & 6,324 & 99.6 & 5 & 133.1& 0.13 & 8.96& 8.96 & 93.9 & 15.2 \\
			& 7.0 & 8,366 & 99.6 & 6 & 177.2 & 0.24 & 8.31 & 8.32 & 91.7 & 12.9 \\
			& 10.0 & 10,000 & 99.7 & 6 & 189.8& -0.17 & 7.82 & 7.82 & 92.7 & 11.4\\
			\specialrule{0em}{3pt}{3pt}
			0.5 & 1.0 & 3,162 & 99.4 & 7 & 143.6 & -0.07 & 9.43 & 9.43 & 92.5 & 20.2 \\
			& 4.0 & 6,324 & 99.6 & 8 & 213.9 & 0.17 & 7.28 &7.28 & 93.4 & 12.1 \\
			& 7.0 & 8,366 & 99.6 & 9 & 266.1 &  0.13 & 6.79 & 6.79 & 93.0 & 10.4\\
			& 10.0 & 10,000 & 99.7 & 10 & 316.6 & 0.24 & 6.00 & 6.01 & 93.2 & 9.0 \\
			
			\specialrule{0em}{1pt}{1pt}
			\hline
		\end{tabular}
	\end{table}

	\begin{table}[!htp]
		
		\caption{ Numerical performance for  estimators $\hat{\gamma}_{\text{AML}}$ corresponding to  simple average  and weighted average schemes.  The generating function is $t(1)$ distribution.
			For imbalanced case, the subsample size $n_k = 1.5n$ for $1 \leq k \leq K/2$ and $n_k = 0.5n$ for $1+K/2 \leq k \leq K$. The simulation experiment is replicated for 1,000 times.} 
		\centering 
		\vspace{0.25 cm}
		\begin{tabular}{rrp{1cm}<{\centering}ccccc}
			\hline
			\hline
			\specialrule{0em}{2pt}{2pt} 
			\multicolumn{1}{c}{\multirow{2}{*}{$N$}} & \multicolumn{1}{c}{\multirow{2}{*}{$n$}}  & \multirow{2}{*}{$K$} & \multirow{2}{*}{Method}& 	Bias & 	SD  &	RMSE \\ 
			\cline{5-7}	
			\\[-2.2em] &&&&&& \\
			&&&& \multicolumn{3}{c}{($\times 10^{-2}$)} 
			\\   \specialrule{0em}{1pt}{1pt}
			\hline
			\specialrule{0em}{3pt}{3pt}
			\multicolumn{7}{c}{{\sc Case 1 (balanced)}}	\\ \specialrule{0em}{3pt}{3pt}
			$1 \times 10^6$ & 1,000 & 10  & \multicolumn{1}{l}{Simple Average}   &  1.12 & 3.89& 4.05 \\
			& & & \multicolumn{1}{l}{Weighted Average} & 1.13 & 3.89 &  4.05 \\
			$5 \times 10^6$ & 2,236 & 14  & \multicolumn{1}{l}{Simple Average}   &  0.46 & 2.50 & 2.54 \\
			& & & \multicolumn{1}{l}{Weighted Average} &  0.47 & 2.49 & 2.53 \\
			$10 \times 10^6$  & 10,000 & 20  & \multicolumn{1}{l}{Simple Average}   &  0.42 & 1.91 & 1.95  \\
			& & & \multicolumn{1}{l}{Weighted Average} &  0.42 & 1.91 &  1.95 \\
			\specialrule{0em}{3pt}{3pt}
			\multicolumn{7}{c}{{\sc Case 2 (imbalanced)}}	\\ \specialrule{0em}{3pt}{3pt}
			$1 \times 10^6$ & 1,000 & 10  & \multicolumn{1}{l}{Simple Average}   &  1.15 & 4.48& 4.62\\
			& & & \multicolumn{1}{l}{Weighted Average} & 1.11 & 3.82 &  3.98 \\
			$5 \times 10^6$ & 2,236 & 14  & \multicolumn{1}{l}{Simple Average}   &  0.67 & 3.20 & 3.27 \\
			& & & \multicolumn{1}{l}{Weighted Average} &  0.64 & 2.69 & 2.76 \\
			$10 \times 10^6$  & 10,000 & 20  & \multicolumn{1}{l}{Simple Average}   &  0.50 & 2.46 & 2.51  \\
			& & & \multicolumn{1}{l}{Weighted Average} &  0.48 & 2.16 &  2.22 \\
			
			\specialrule{0em}{1pt}{1pt}
			\hline
		\end{tabular}
		\label{tabX}	
	\end{table}

	\begin{table}[t]
		
		\caption{ The $\hat{\gamma}_{\text{AML}}$, estimated upper bound (99.99\% quantile) and the correlated suspected outlier probability (Sop) for four continuous variables we focused, based on $K=100$ subsamples of size $n = 10,000$.  The kurtosis (Kurt) and the $\hat{\gamma}_{\text{global}}$  based on the whole dataset for each variable is also reported.} 
		\centering 
		\vspace{0.25 cm}
		\begin{tabular}{lrrrrr}
			\hline
			\hline
			\specialrule{0em}{2pt}{2pt}
			Variable &{Kurt}&  $\hat{\gamma}_{\text{global}}$ & $\hat{\gamma}_{\text{AML}}$  & \multicolumn{1}{c}{Upper Bound}&Sop(\%)\\ \cline{1-6}
			\specialrule{0em}{2pt}{2pt}
			ActualElapsedTime & 5.6 & 0.1176 & 0.1165 & 583.7 & 0.007 \\
			CRSElapsedTime & 5.7 &  0.1127 & 0.1123 & 573.1 & 0.007\\
			ArrDelay& 55.3 & 0.2881 & 0.2874 &  525.0 & 0.009\\
			DepDelay& 275.3 & 0.3118 & 0.3129 &565.7 & 0.011\\
			\specialrule{0em}{1pt}{1pt}
			\hline
		\end{tabular}
		\label{tab_real}
	\end{table}
	
	\def\fillandplacepagenumber{%
		\par\pagestyle{empty}%
		\vbox to 0pt{\vss}\vfill
		\vbox to 35pt{\baselineskip0pt
			\hbox to\linewidth{\hss}%
			\baselineskip\footskip
			\hbox to\linewidth{%
				\hfil\thepage\hfil}\vss}}
	
	\newpage
	\begin{landscape}
		\begin{center}
			\begin{figure}[h]
				\centering
				\includegraphics[scale=.5]{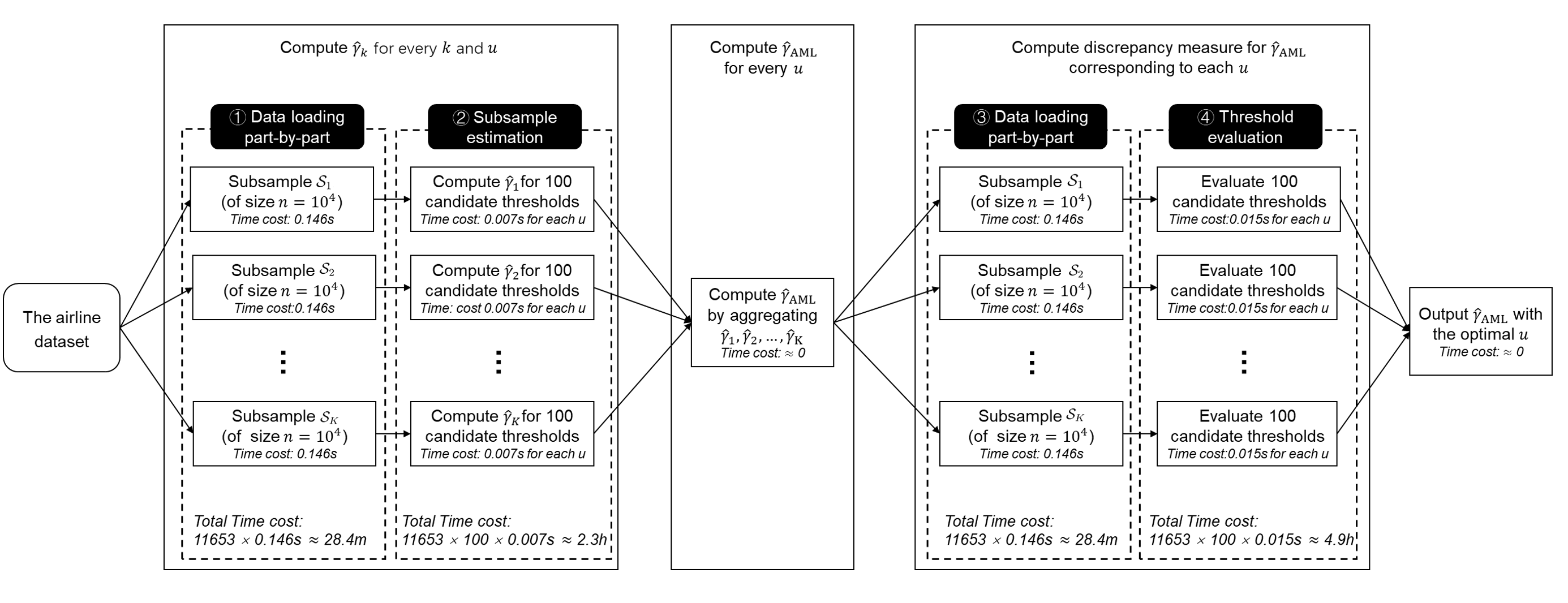}
				\caption{Time cost for each step of the airline dataset analysis. The total time cost for processing the whole dataset $\approx  28.4m \times 2 + 2.3h + 4.9h \approx 8$ hours. If subsampling method with $K=100$ is used, then the total time cost become $(0.146 \times 2 + 100 \times 0.007s + 100 \times 0.015s) \times 100 \approx 4.2 $ minutes. All analyses are carried out on a Ubuntu 18.04 desktop computer with a 2.2GHz intel Xeon Silver 4210 processor by using Python 3.7. 
				} 
				\label{workflow} 
			\end{figure}
		\end{center}
		\fillandplacepagenumber
	\end{landscape}

\end{document}